\documentclass[pra,aps,superscriptaddress,notitlepage,longbibliography,twocolumn,nofootinbib,superscriptaddress]{revtex4-2}
\usepackage{amsmath}
\usepackage{amsfonts}
\usepackage{graphicx}
\usepackage{bm}
\usepackage{bbold}
\usepackage{color}
\usepackage{braket}
\usepackage{tikz}
\usetikzlibrary{positioning}
\usepackage{comment}

\usepackage{xcolor}

\usepackage[colorlinks]{hyperref}

\hypersetup{
    colorlinks=true,
    linkcolor=blue,
    filecolor=magenta,      
    urlcolor=magenta,
    citecolor={blue},
    }

\begin{document}

\abovedisplayskip=6pt
\abovedisplayshortskip=6pt
\belowdisplayskip=6pt
\belowdisplayshortskip=6pt

\title{Impact of mode regularization for quasinormal mode perturbation theories}
\author{
Sebastian Franke}
\email{sebastian.r.franke@gmail.com}
\affiliation{Technische Universit\"at Berlin, Institut f\"ur Theoretische Physik,
Nichtlineare Optik und Quantenelektronik, Hardenbergstra{\ss}e 36, 10623 Berlin, Germany}
\affiliation{\hspace{0pt}Department of Physics, Engineering Physics, and Astronomy, Queen's University, Kingston, Ontario K7L 3N6, Canada\hspace{0pt}}
\author{Juanjuan Ren}
\affiliation{\hspace{0pt}Department of Physics, Engineering Physics, and Astronomy, Queen's University, Kingston, Ontario K7L 3N6, Canada\hspace{0pt}}
 \author{Stephen Hughes}
\affiliation{\hspace{0pt}Department of Physics, Engineering Physics, and Astronomy, Queen's University, Kingston, Ontario K7L 3N6, Canada\hspace{0pt}}

\begin{abstract}
We give insight into the critical problem of an open resonator that is subject to a  perturbation outside of its cavity region. We utilize the framework of  quasinormal modes (QNMs), which are the natural mode solutions to the open boundary problem with complex eigenfrequencies. We first highlight some fundamental problems with currently adopted formulas
using QNM perturbation theory, when perturbations are added outside the resonator structure and present a first potential step for solving this problem connected to a regularization of the QNMs. 
We then show an example for a full three-dimensional plasmonic resonator of arbitrary shape and complex dispersion and loss, clearly displaying the divergent nature of the first-order mode change predicted from QNM perturbation theory.
Subsequently, we concentrate on the illustrative case of a one-dimensional dielectric barrier, where analytical QNM solutions are possible. We inspect the change of the mode frequency as function of distance between the cavity and another smaller barrier structure. The results obtained from a few QNM expansion are compared with exact analytical solutions from a transfer matrix approach.
We show explicitly how regularization prevents a problematic spatial divergence for QNM perturbations in the far field, though eventually higher-order effects and multimodes can also play a role in the full scattering solution, and 
retaining a pure discrete QNM picture becomes questionable in such situations, since the input-output coupling ultimately involves reservoir modes. 
\end{abstract}

\maketitle
\section{Introduction}
Perturbation theory of open resonators is an  important topic in quantum and classical optics and has a variety of applications, such as detection and sensing in the vicinity of a scattering object, including plasmonic or dielectric cavities 
\cite{Vyas2008,Waldron1960,imanec1971,Lalanne_review}. The introduction of a perturbation into a resonator environment, e.g., a metallic tip, and its change of the permittivity functions leads to a change of the resonator's eigenmodes and frequencies. Alternative ways of perturbing the system include the deformation of the scattering object or the overall change of the material.

There are several theoretical techniques to describe these changes in the cavity mode properties, which are usually based on so-called modes of the universe~\cite{motu1,motu2,Huttner}, i.e., solutions of the Helmholtz equation with vanishing boundary conditions at $|\mathbf{r}|\rightarrow \infty$, or, in an approximate model, using so-called normal modes (NMs), which are solutions of the Helmholtz equation with fixed boundary conditions of the associated closed resonator~\cite{GirishBook1}. While the former case involves a continuous set of modes, which is not tractable in numerical simulations, the latter approach is only reliable for very high quality factors, namely, small radiation leakage of the resonators. Indeed,  the NM approach is only rigorous in the case of no loss, since the eigenfrequencies are real, and is thus generally ambiguous for open cavities, even for high $Q$ (quality factor)
resonators.

A rigorous open resonator method not only allows one to investigate a change in the oscillation frequency of the cavity mode, but also a change of its spectral linewidth and temporal decay, which is not at all possible using a NM description (real frequencies). The general eigensolution to this open cavity problem are so-called {\it quasinormal modes}~\cite{Lai,LeungSP1D,2ndquant2,KristensenHughes,NormKristHughes,Kristensen:20,Lalanne_review} (QNMs), which have outgoing boundary conditions and complex mode frequencies. The QNMs can also be formulated for general dispersive and absorptive cavities, where even the modes of the universe description in its usual form cannot be used anymore~\cite{Leung3}. 

Various QNM and open cavity modes techniques have been used quite extensively in recent years, for a successful description of light-matter interactions close or within the resonator object~\cite{muljarovPert,KristensenHughes,SauvanNorm,NormKristHughes,PhysRevA.98.043806,Lalanne_review,carlson2019dissipative,PhysRevA.98.043806,PhysRevB.100.155406}. 
This also includes a generalization of well-known perturbation theories, which have been typically described on the basis of NMs. In particular, this allows one to describe perturbation inside an open cavity~\cite{Kristensen:14} or even close to an open cavity region for near field optics in the linear~\cite{Giessen15,Cognee:19} as well as non-linear regime~\cite{Christopoulos:20}.

\begin{figure}[h]
    \centering
    \includegraphics[width=0.99\columnwidth]{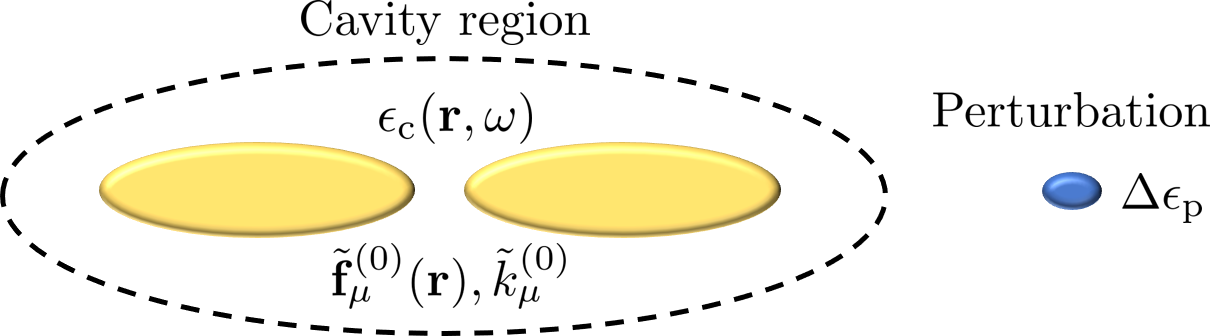}
    \caption{Schematic of a resonator with permittivity $\epsilon_{\rm c}(\mathbf{r},\omega)$, 
    perturbed by a small source with permivitity difference $\Delta\epsilon_{\rm p}$  
    outside of the cavity region. The unperturbed cavity supports a set of QNMs with eigenfunctions $\tilde{\mathbf{f}}^{(0)}_\mu(\mathbf{r})$ and eigen wavenumbers $\tilde{k}_\mu^{(0)}=\tilde{\omega}_\mu^{(0)}/c$, which form a complete representation of the field inside the cavity region.
}\label{fig: Schematic}
\end{figure}

However, the spatial divergence of the QNMs prevents one to describing phenomena that happen far outside the resonator region, as depicted in Fig.~\ref{fig: Schematic}. The divergence stems from the open boundary conditions in combination with complex eigenfrequencies with a negative imaginary part. Importantly, this is not just a particular problem of QNMs, but is an intrinsic property of open cavity systems, that are not phenomenologically treated as closed systems with dissipation added {\it ad-hoc}. In terms of perturbation theory, this ``problem'' has not been addressed in the literature and, to our knowledge, only partly highlighted recently in Refs.~\onlinecite{PhysRevX.11.041020,Christopoulos:20}.

While much progress has been made into rigorously fixing the divergent behavior of QNMs for formulating a well-defined Purcell factor~\cite{GeNJP2014,RegQNMs} (i.e., the enhanced/suppressed spontaneous emission rate of a dipole emitter), general far-field quantities~\cite{PhysRevB.102.035432}, and a rigorous QNM quantization method~\cite{PhysRevLett.122.213901,franke2020fluctuation,PhysRevB.101.205402}, not much has been done in terms of a generalized perturbation theory. Similar to using a closed cavity approach, the QNM theories can start to become invalid in this regime, and are generally ambiguous for any perturbation {\it outside} the cavity structure. In a more general view, the same conceptional problem applies to practical cavity circuits, such as waveguide-cavity systems, where two well-separated open and lossy resonators may interact through a waveguide structure (involving the propagating waveguide modes). As recognized and discussed recently~\cite{Yu:21}, this is a promising structure for setting up a quantum network of several quantum emitters attached to the cavities.

In this work, we highlight the intrinsic problem with QNM perturbation theory and discuss methods to potentially fix them.
In Sec.~\ref{Sec: QNMs}, we introduce some essential background QNM theory,
and discuss a Green function expansion in terms of QNMs and regularized QNMs.
In Sec.~\ref{Sec:pt}, we present the perturbation theory for QNMs and highlight the general modelling problem with adding perturbations outside the cavity region, which renders current approaches ambiguous in general.
In Sec.~\ref{sec: App}, we then present several applications for the one-dimensional case, including the calculation of the local density of states (LDOS) for a single-barrier cavity as well as the calculation of eigenfrequencies of a double-barrier cavity system, using an exact transfer matrix approach and a first-order perturbation theory. In Sec.~\ref{Sec: Discussions}, we then discuss the whole problem from different views, and also elaborate on a possible alternative representation of the QNM Green function (which is non-diagonal in the QNM expansion).
Finally, in Sec.~\ref{Sec: Conclusions}, we give our conclusions.

In addition,
we also present several appendices, including the equivalence of different regularized QNM approaches, the derivation of two Green identities as well as details of the contour integration in connection with a non-diagonal form of a QNM Green function.

\section{Quasinormal mode theory\label{Sec: QNMs}}

In this section, we  briefly present the essential background theory and properties of QNMs in the language of perturbation theory. First, we look at the isolated open cavity structure embedded in a lossless background.

The unperturbed QNM eigenfunctions, $\tilde{\mathbf{f}}_\mu^{(0)}$, are solutions to the Helmholtz equation 
\begin{equation}
    \left\{\boldsymbol{\nabla}\times\boldsymbol{\nabla}\times-\left[\tilde{k}_\mu^{(0)}\right]^2\epsilon_{\rm c}(\mathbf{r},\tilde{\omega}_\mu^{(0)})\right\}\tilde{\mathbf{f}}_\mu^{(0)}(\mathbf{r})=0,\label{eq: HelmholtzQNM}
\end{equation}
together with open boundary conditions, i.e., the Silver-Müller radiations conditions, 
\begin{equation}
    \frac{\mathbf{r}}{|\mathbf{r}|}\times\boldsymbol{\nabla}\times\tilde{\mathbf{f}}_\mu^{(0)}(\mathbf{r})\rightarrow in_{\rm B}\tilde{k}_\mu^{(0)} \tilde{\mathbf{f}}_\mu(\mathbf{r}),\label{eq: SM_cond}
\end{equation}
which are asymptotic relations for $|\mathbf{r}|\rightarrow \infty$. Here, $\tilde{k}_\mu^{(0)}=\tilde{\omega}_\mu^{(0)}/c$, where $c$ is the vacuum speed of light, and $\tilde{\omega}_\mu^{(0)}=\omega_\mu^{(0)}-i\gamma_\mu^{(0)}$ is the unperturbed QNM eigenfrequency, which is a complex number due to the open boundary conditions. In the following, we use the term ``eigenfrequency'' for $\tilde{k}_\mu$ and $\tilde{\omega}_\mu$ interchangeably (implicitly setting $c=1$). 

The associated QNM resonance has a real part of the frequency $\omega_\mu^{(0)}$ and a half width at half maximum $\gamma_\mu^{(0)}$ (which comes from the imaginary part of the eigenfrequency, where $\gamma_\mu^{(0)}=\omega_\mu^{(0)}/[2Q_{\mu}^{(0)}]$). Furthermore, $\epsilon_{\rm c}(\mathbf{r},\tilde{\omega}_\mu^{(0)})$ is the analytical continuation of the complex-valued permittivity function $\epsilon_{\rm c}(\mathbf{r},\omega)$, that  describes a spatial-inhomogeneous and (possibly) dispersive resonator geometry embedded in a background medium, with constant and real-valued refractive index $n_{\rm B}$, as visualized in Fig.~\ref{fig: Schematic} (shown without an additional perturbation). 

When properly normalized, the QNMs can be used as a basis to expand the transverse part of the Green function $\mathbf{G}(\mathbf{r},\mathbf{r}_0,\omega)$, which describes the light propagation from a source point $\mathbf{r}_0$ to $\mathbf{r}$, and is formally defined via the Helmholtz equation 
\begin{equation}
     [\boldsymbol{\nabla}\times\boldsymbol{\nabla}\times-k_0^2\epsilon_{\rm c}(\mathbf{r},\omega)]\mathbf{G}(\mathbf{r},\mathbf{r}_0,\omega)=k_0^2\delta(\mathbf{r}-\mathbf{r}_0),\label{eq: GreenHelmholtz}
\end{equation}
together with suitable radiation conditions, namely Eq.~\eqref{eq: SM_cond}, for real-valued frequencies $\omega$ and $k_0=\omega/c$. The resulting QNM expanded form of the transverse part of the Green function, $\mathbf{G}^{\perp}(\mathbf{r},\mathbf{r}_0,\omega)$, for spatial positions inside the resonator geometry, reads
\begin{equation}
    \mathbf{G}^{\perp}(\mathbf{r},\mathbf{r}_0,\omega)=\sum_\mu A_\mu^{(0)}(\omega)\tilde{\mathbf{f}}_\mu^{(0)}(\mathbf{r})\tilde{\mathbf{f}}_\mu^{(0)}(\mathbf{r}_0)\label{eq: GreenQNM},
\end{equation}
with $A_\mu^{(0)}(\omega)=\omega/(2(\tilde{\omega}_\mu^{(0)}-\omega))$, and where, in principle, the sum runs over all $\mu=0,\pm 1, \pm 2,\dots$ with the imposed ordering $0<\omega_1<\omega_2<\dots$ and symmetry properties $\tilde{\mathbf{f}}_{-\mu}^{(0)}=\tilde{\mathbf{f}}_{\mu}^{(0)*}$ as well as $\tilde{\omega}_{-\mu}^{(0)}=-\tilde{\omega}_\mu^{(0)*}$. We note, that although the full set of modes including the so-called ``zero mode''~\cite{leung1997two} ($\mu=0$) are required for general completeness of QNMs, only few QNMs with $\mu>0$ are needed in practical situations of classical and quantum optics.

One of the drastic consequences of the open boundary conditions, is that the QNMs spatially diverge for far field positions, and completeness can only be achieved within the resonator region. Strictly speaking, this is correct only for resonator embedded in a spatial-homogeneous background and completeness was only proven explicitly for certain resonator types, e.g., of spherical shape or one-dimensional cavities~\cite{LeungSP1D,Leung3,MDR1}. 
In general, for the common case of resonators lying
on substrates, branch cuts exist and the QNM basis is incomplete, cf.~Refs.~\onlinecite{Lalanne_review,Kristensen:20} and references
therein. Completeness is restored in advanced theories using additional numerical modes, see Refs.~\onlinecite{PhysRevA.89.023829,PhysRevB.97.205422} for instance.
Since completeness is required for an eigenfunction expansion of the Green function, Eq.~\eqref{eq: GreenQNM} must be adapted for positions in the background region. Otherwise, this would prevent one from using the QNM expansion as a basis to calculate any spatial overlap outside of the cavity region, which is necessary to calculate frequency changes for perturbations located in the background region.

It is noteworthy that, for the case of purely amplifying media and where gain overcompensates the radiative loss, the imaginary part of the QNM eigenfrequencies would change the sign of $\gamma_\mu$. As a consequence, the resulting {\it gain} QNMs~\cite{PhysRevX.11.041020} would be spatially damped in the far field, and the divergence would instead appear in the time domain. However, treating such cases on the level of linear amplification would not be consistent with the causality relation or (in frequency space) the Kramers-Kronig relations of $\mathbf{G}(\mathbf{r},\mathbf{r}_0,\omega)$, which would be necessary to formulate a QNM quantization formalism~\cite{PhysRevLett.127.013602,PhysRevA.105.023702}.

As an option to circumvent or resolve this QNM  divergence problem, one can introduce a {\it regularization} procedure of the QNMs by exploiting an integral form of the Green function or the field equivalence principle, where the former is related to the well-known (classical) Dyson equation. Such regularization not only prevents any unphysical behavior of important classical quantities, such as an increasing enhancement of the spontaneous emission in the far field of a metallic antenna~\cite{GeNJP2014,RegQNMs}, but it is also required to properly predict the radiative dissipation in a quantized QNM theory~\cite{PhysRevLett.122.213901,franke2020fluctuation,PhysRevB.101.205402}. 

Furthermore, it is important to note that this is not just a heuristic fix for diverging QNMs (which is sometimes misunderstood in the literature), and the resulting regularized QNM functions are rigorously defined, in that only fundamental Green identities as well as the completeness of the QNMs {\it inside} the resonator volume (or at the resonator boundary) are required. These QNM functions carry over the characteristics of the (infinitely extended) background medium through a real continuous frequency $\omega$ for each discrete QNM, which is physically intuitive and in line with the system-bath and input-output theories, where they can be interpreted as reservoir modes. We elaborate more on that point in this and following sections.
Indeed, not using these regularized QNMs would be problematic for a certain class of problems if one wants to keep with a few mode description.

Next, we briefly elaborate and shed  further insights on the different QNM regularization procedures. All the methods exploit the assumed completeness of the QNMs. As a consequence of the this completeness, the full (transverse) electric field $\mathbf{E}(\mathbf{r},t)$ (of the unperturbed structure) at position $\mathbf{r}$ in the cavity region can be represented as
\begin{equation}
    \mathbf{E}(\mathbf{r},t)=\sum_\mu a_\mu^{(0)}(t)\tilde{\mathbf{f}}_\mu^{(0)}(\mathbf{r}),
\end{equation}
where $a_\mu^{(0)}(t)=e^{-i\tilde{\omega}_\mu^{(0)} t}a_\mu^{(0)}(t=0)$ is the harmonic solution to the temporal part of the wave equations.

With respect to the former approach, a regularized QNM can be defined through formulating the analogue of Eq.~\eqref{eq: HelmholtzQNM} for the Fourier transform of $ \mathbf{E}(\mathbf{r},t)$, namely $\boldsymbol{\mathcal{E}}(\mathbf{r},\omega)$, as a scattering problem. This leads to the following solution:
\begin{align}
    \boldsymbol{\mathcal{E}}(\mathbf{R},\omega)=&\boldsymbol{\mathcal{E}}_{\rm hom}(\mathbf{R},\omega)\nonumber\\
    &+\int{\rm d}^3r\Delta\epsilon_{\rm c}(\mathbf{r},\omega)\mathbf{G}_{\rm B}(\mathbf{R},\mathbf{r},\omega)\cdot\boldsymbol{\mathcal{E}}(\mathbf{r},\omega)\label{eq: DysonRegF},
\end{align}
where $\mathbf{G}_{\rm B}(\mathbf{R},\mathbf{r},\omega)$ is the background Green function, i.e., the solution of Eq.~\eqref{eq: GreenHelmholtz} with $\epsilon_{\rm c}(\mathbf{r},\omega)\rightarrow n_{\rm B}^2$, and $\Delta\epsilon_{\rm c}(\mathbf{r},\omega)=\epsilon_{\rm c}(\mathbf{r},\omega)-n_{\rm B}^2$ restricts the spatial integral to the cavity volume.
Since we demand consistency with the Silver-Müller radiations conditions, the only possible homogeneous solution is $\boldsymbol{\mathcal{E}}_{\rm hom}(\mathbf{R},\omega)=\mathbf{0}$. 

Using the completeness relation of the QNMs inside the resonator volume, we can rewrite Eq.~\eqref{eq: DysonRegF} as
\begin{equation}
    \boldsymbol{\mathcal{E}}(\mathbf{R},\omega)=\sum_\mu  a_\mu^{(0)}(\omega)\tilde{\mathbf{F}}_\mu^{(0)}(\mathbf{R},\omega), \label{eq: EFour_vol}
\end{equation}
where $a_\mu^{(0)}(\omega)=2\pi i a_\mu^{(0)}(t=0)/(\tilde{\omega}_\mu^{(0)}-\omega)$, and 
\begin{equation}
    \tilde{\mathbf{F}}_\mu^{(0)}(\mathbf{R},\omega)=\int{\rm d}^3r\Delta\epsilon_{\rm c}(\mathbf{r},\omega)\mathbf{G}_{\rm B}(\mathbf{R},\mathbf{r},\omega)\cdot\tilde{\mathbf{f}}_\mu^{(0)}(\mathbf{r})\label{eq: DysonRegF2}.
\end{equation}

The function, $\tilde{\mathbf{F}}_\mu^{(0)}(\mathbf{r},\omega)$, is a function of {\it real frequency} $\omega$. The QNM eigenfunctions $\tilde{\mathbf{f}}_\mu^{(0)}(\mathbf{r})$ are fundamentally connected to the above regularized QNM functions via analytical continuation of $\omega$ into the complex area, i.e., $\tilde{\mathbf{F}}_\mu^{(0)}(\mathbf{r},\tilde{\omega}_\mu^{(0)})=\tilde{\mathbf{f}}_\mu^{(0)}(\mathbf{r})$, and thus we also refer to $\tilde{{\bf F}}_\mu$ as a mode, but one which is now regularized. 
Note, that in the one-dimensional case for a cavity with boundaries $x=a,b$, there is a rather interesting relation between $\tilde{F}_\mu(x>b,\omega_\mu)$ and $\tilde{F}_\mu(x>b,\tilde{\omega}_\mu)$, namely,
\begin{equation}
    \tilde{F}_\mu(x,\omega_\mu)=e^{-\gamma_\mu n_{\rm B}(x-b)/c}\tilde{F}_\mu(x,\tilde{\omega}_\mu),
\end{equation}
which will be discussed in more detail in Sec.~\ref{sec: App}. 

In the more general three-dimensional case, a similar relation does only exists for $|\mathbf{r}|\rightarrow\infty$ through 
\begin{equation}
    \tilde{\mathbf{F}}_\mu(\mathbf{r},\omega_\mu)\rightarrow e^{-\gamma_\mu n_{\rm B}|\mathbf{r}|/c}\tilde{\mathbf{F}}_\mu(\mathbf{r},\tilde{\omega}_\mu),
\end{equation}
which is still an approximate relation, since $\tilde{\mathbf{F}}_\mu(\mathbf{r},\omega)$ contains additional $\omega$-dependent terms, that are not part of the exponential function.
If such a relation would exist for all positions outside the resonator, one could significantly improve the numerical calculations, as one would only need the QNM for outside positions multiplied by $e^{-\gamma_\mu n_{\rm B}|\mathbf{r}|/c}$.

Exploiting a form of Green's theorem together with the Helmholtz equation of $\boldsymbol{\mathcal{E}}(\mathbf{r},\omega)$ and $\mathbf{G}_{\rm B}(\mathbf{R},\mathbf{r},\omega)$, one can find an alternative expression for the Fourier transform $\boldsymbol{\mathcal{E}}(\mathbf{R},\omega)$ outside the resonator region 
\begin{equation}
    \boldsymbol{\mathcal{E}}(\mathbf{R},\omega)=\sum_\mu  a_\mu^{(0)}(\omega)\tilde{\mathbf{F}}_\mu^{\prime(0)}(\mathbf{R},\omega), \label{eq: EFour_sur}
\end{equation}
with
\begin{align}
    \tilde{\mathbf{F}}_\mu^{\prime(0)}&(\mathbf{R},\omega)=\frac{c^2}{\omega^2}\int_{\mathcal{S}'}{\rm d}A_{\mathbf{s}} \mathbf{G}_{\rm B}(\mathbf{R},\mathbf{s},\omega)\cdot[\mathbf{n}\times\boldsymbol{\nabla}\times\tilde{\mathbf{f}}_\mu^{(0)}(\mathbf{s})]\nonumber\\
    &-\frac{c^2}{\omega^2}\int_{\mathcal{S}'}{\rm d}A_{\mathbf{s}}[\mathbf{n}\times\boldsymbol{\nabla}\times\mathbf{G}_{\rm B}(\mathbf{s},\mathbf{R},\omega)]^{\rm t}\cdot\tilde{\mathbf{f}}_\mu^{(0)}(\mathbf{s}), \label{eq: FregFEP}
\end{align}
where
$\mathcal{S}'$ is a surface that surrounds the scattering volume defined by $\Delta\epsilon_{\rm c}(\mathbf{r},\omega)$, $\mathbf{R}$ is located outside of $\mathcal{S}'$ and the superscript `t' denotes the transpose. We note, that an implicit condition for this form is assuming that the QNMs are a valid representation for the electric field at $\mathcal{S}'$, and thus one is usually limited to surfaces very close to the scattering volume.
We further note that this result can also be obtained via the field equivalence principle~\cite{schelkunoff1936some} or similarly by a near-field to far-field transformation technique~\cite{PhysRevB.101.205402}. Choosing $\mathcal{S}'$ as the surface of the scattering volume leads to the equivalence of Eq.~\eqref{eq: EFour_sur} and Eq.~\eqref{eq: EFour_vol}, as shown in App.~\ref{app: SurVolRep}.  
However, these representations are not identical in a few-QNM approximation, since different parts of the QNM set will contribute differently to the overall completeness at $\mathcal{S}'$ and in $V$ (scattering volume). Nevertheless, as shown in Ref.~\onlinecite{PhysRevB.101.205402}, both regularization behave similarly for far field positions, i.e., for $|\mathbf{R}|\gg \lambda_\mu^{(0)}$, where $\lambda_\mu^{(0)}=2\pi c/(n_{\rm B}\omega_\mu^{(0)})$ is the associated wavelength of the unperturbed QNM.

\section{Perturbation theory of open resonators}
\label{Sec:pt}
In Section~\ref{Sec: QNMs}, we have discussed the unperturbed problem with a single cavity supporting a set of QNMs with eigenfunctions and eigenfrequnencies $\{\tilde{\mathbf{f}}_\mu^{(0)},\tilde{\omega}_\mu^{(0)}\}$. When adding another
scattering object with volume $V$ and permittivity $\epsilon_{\rm p}(\mathbf{r},\omega)=\epsilon_{\rm B}+\Delta\epsilon_{\rm p}(\mathbf{r},\omega)$, such as a molecule or detector object outside the cavity structures, the permittivty function changes to $\epsilon(\mathbf{r},\omega)=\epsilon_{\rm c}(\mathbf{r},\omega)+\Delta\epsilon_{\rm p}(\mathbf{r},\omega)$. One could again solve the source-free Helmholtz equation by replacing $\epsilon_{\rm c}\rightarrow \epsilon$ to obtain the full QNM eigenfunctions and eigenfrequencies $\{\tilde{\mathbf{f}}_\mu,\tilde{\omega}_\mu\}$. However,  usually, this is a very numerically demanding task, especially for three-dimensional structures. It also offers no analytical insight.

In this section, we will discuss the possibility to utilize the regularization of QNMs for formulating a QNM framework, where the cavity-perturbation setup as visualized in Fig.~\ref{fig: Schematic} could be tackled within the initial unperturbed QNM basis.

First, we will analyze the problem that appears when simply using $\tilde{\mathbf{f}}_\mu^{(0)}(\mathbf{r})$. Adopting the one-dimensional time-independent perturbation technique from Refs.~\onlinecite{Leung2,Leung3}, for a general three-dimensional system, a first-order correction to $\tilde{k}^{(0)}_\mu$ can be calculated in the general dispersive case as 
\begin{align}
    \tilde{k}_\mu^{(1)}&=-\frac{\tilde{k}_\mu^{(0)}}{2}\int{\rm d}^3r\Delta\epsilon_{\rm p}(\mathbf{r},\tilde{\omega}_\mu^{(0)})~\tilde{\mathbf{f}}^{(0)}_\mu(\mathbf{r})\cdot\tilde{\mathbf{f}}^{(0)}_\mu(\mathbf{r}).\label{eq: FirstOrder_3D}
\end{align}
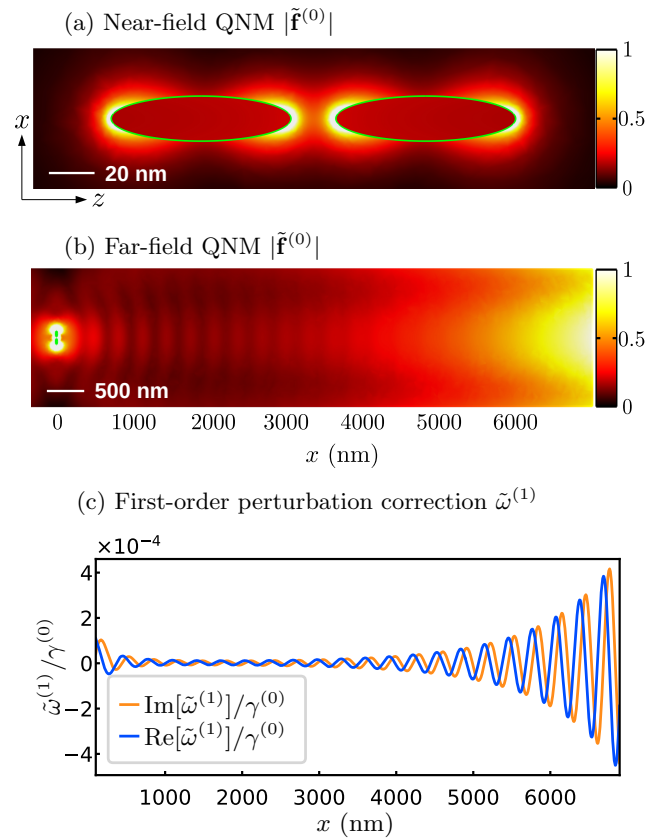
\begin{figure}[t]
    \begin{tikzpicture}
    \centering
        [inner sep=0mm]
\node [xshift=-30mm] (figa) 
{\includegraphics[width=0.99\columnwidth,trim=-1mm 0 0 0cm]{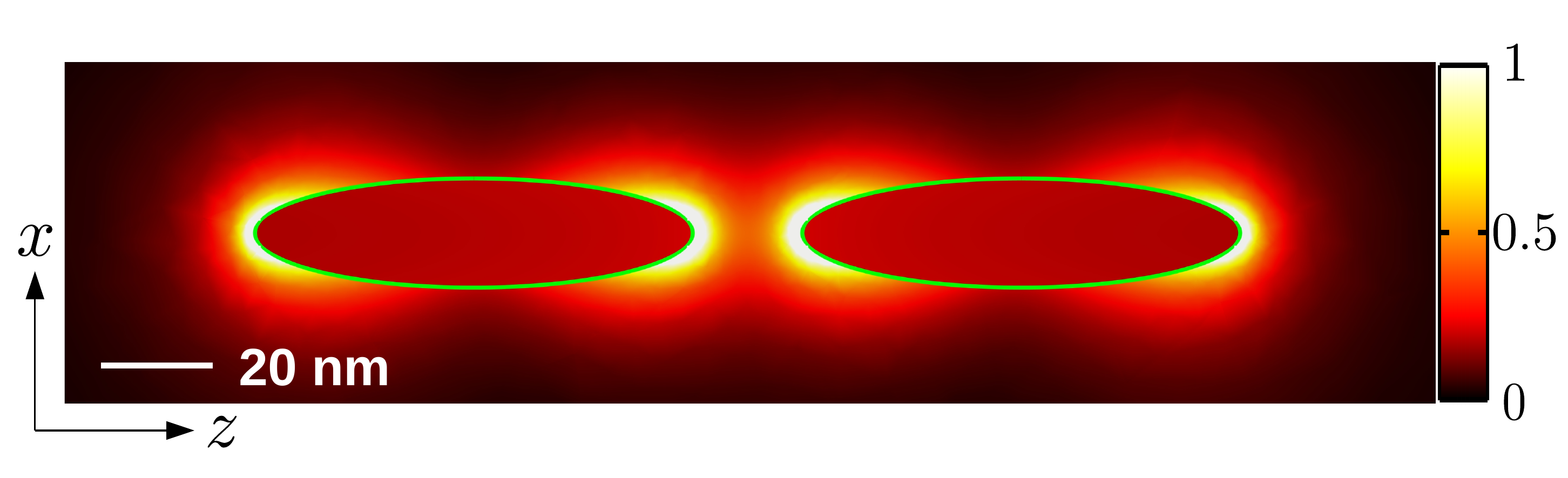}};
\node[below=of figa,xshift=-0mm,yshift=7mm] (figb) {\includegraphics[width=0.99\columnwidth]{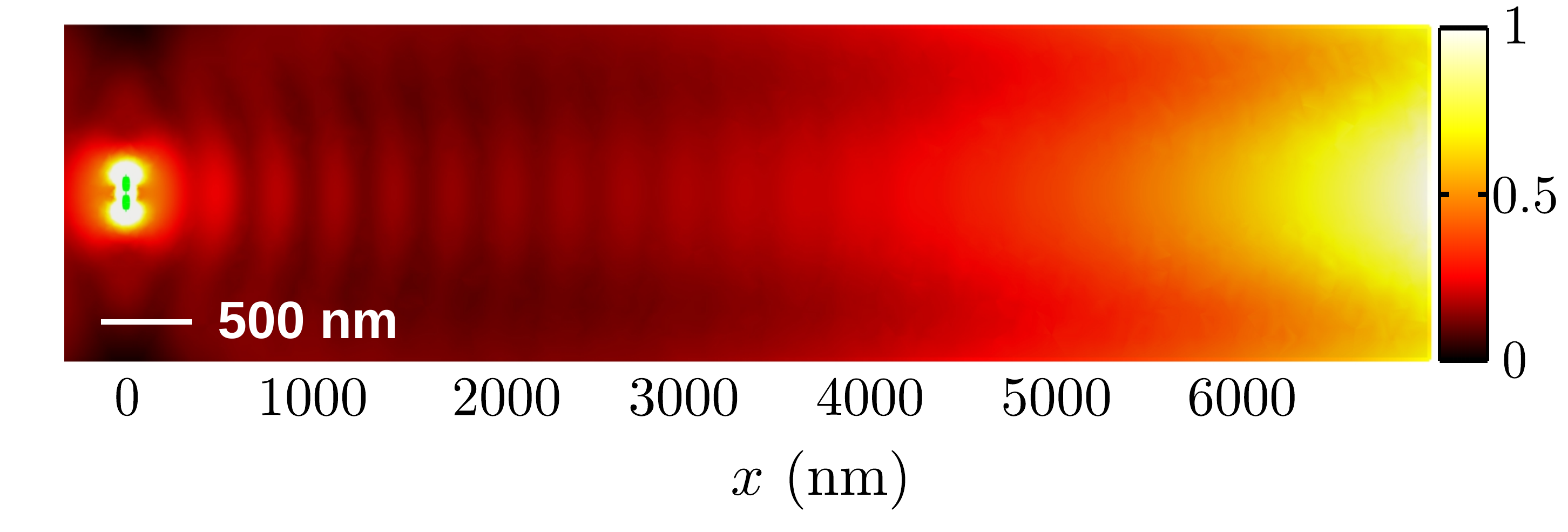}};
\node[below=of figb, xshift=-0mm,yshift=7mm] (figc) 
{\includegraphics[width=0.96\columnwidth]{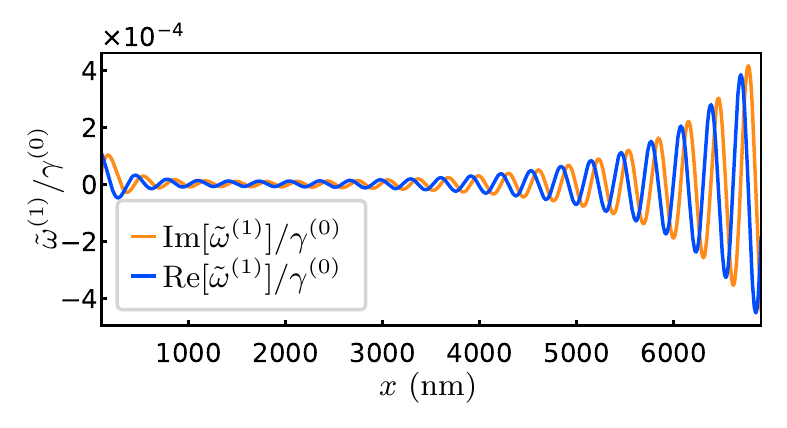}};
\node[xshift=-61mm, yshift=-1mm] at (figa.north east) {(a) Near-field QNM $|\tilde{\mathbf{f}}^{(0)}|$};
\node[xshift=26mm, yshift=+0.6mm] at (figb.north west) {(b) Far-field QNM $|\tilde{\mathbf{f}}^{(0)}|$};
\node[xshift=40mm, yshift=+1mm] at (figc.north west) {(c) First-order perturbation correction $\tilde{\omega}^{(1)}$};
\end{tikzpicture}
\vspace{-0.2cm}
\caption{(a) Near-field QNM (arb. units) $|\tilde{\mathbf{f}}^{(0)}|$ for a gold ellipsoid dimer with gap distance of $20~$nm, as visualized schematically in Fig.~\ref{fig: Schematic}. The center width and length of the ellipsoid are $20~$nm and $80~$nm, respectively. 
Note that the $z$-component is the dominant
field component here, though all three components are included in $|\tilde{\mathbf{f}}^{(0)}|$. (b) Far-field QNM $|\tilde{\mathbf{f}}^{(0)}|$ of the same dimer (also in arb. units). The spatially diverging behavior is clearly shown in the far field region,  starting from around $3000~$nm ($\sim 5\lambda^{(0)}$ the current low $Q$ example) away. (c) First-order perturbation correction $\tilde{\omega}^{(1)}$ from $\tilde{k}^{(1)}$ [first line of Eq.~\eqref{eq: k1_fR}]. The perturbation region is a sphere with radius of $10~$ nm and the dielectric constant is $\epsilon_{\rm p}=4.8$ ($\Delta\epsilon_{\rm p}=\epsilon_{\rm p}-\epsilon_{\rm B}=3.8$). $\tilde{\omega}^{(0)}=\omega^{(0)}-i\gamma^{(0)}$ is the unperturbed resonance frequency of the dimer.  
}
\label{fig: QNMSingle_dimer}
\end{figure}

To show the conceptional failure of the QNMs in a very simple and intuitive way,
we can  move a point-like perturbation to the very far field at $\mathbf{X}_{\rm far}$ and then derive the first-order perturbation modification $\tilde{k}^{(1)}_\mu$ 
as
\begin{align}\label{eq: k1_fR}
    \tilde{k}_\mu^{(1)}&=-\frac{\tilde{k}_\mu^{(0)}}{2}V_{\rm p}\Delta\epsilon_{\rm p}~\tilde{\mathbf{f}}^{(0)}_\mu(\mathbf{X}_{\rm far})\cdot\tilde{\mathbf{f}}^{(0)}_\mu(\mathbf{X}_{\rm far})\nonumber\\
    &\propto -\frac{\tilde{k}_\mu^{(0)}}{2}V_{\rm p}\Delta\epsilon_{\rm p}~e^{in_{\rm B}\omega_\mu^{(0)} |\mathbf{X}_{\rm far}|/c}e^{n_{\rm B}\gamma_\mu^{(0)} |\mathbf{X}_{\rm far}|/c},
\end{align}
whose form is a consequence of the Silver-Müller radiation conditions.
Although this is the currently accepted form
of QNM perturbation theory, 
it certainly diverges for $|\mathbf{X}_{\rm far}|\rightarrow \infty$. Here, $V_{\rm p}$ is the point-like perturbation volume.  
In a more quantitative manner, we also show $\tilde{k}^{(1)}_\mu$ 
for the structure depicted in Fig.~\ref{fig: QNMSingle_dimer} (a), where we move a small spherical volume with radius $r_{\rm p}=10~{\rm nm}$ and permittivity difference $\Delta\epsilon_{\rm p}=3.8$ ($n_{\rm B}=1$) from a near-field region of a plasmonic dimer to the far-field region along the $x$ direction. To model the permittvity of the plasmonic dimer, we use the spatial-homogeneous Drude-model, 
\begin{equation}
    \epsilon(\omega)=1-\frac{\omega_{\rm p}^2}{\omega(\omega+i\gamma_{\rm p})},
\end{equation}
with parameters that are somewhat similar to gold, with a plasma frequency $\omega_{\rm p}=\omega_{\rm p,au}=1.26\times10^{16}$~(rad/s) and $\gamma_{\rm p}=3\gamma_{\rm p,au}=4.23\times10^{14}$~(rad/s). An efficient dipole-scattering approach is used to get the unperturbed QNM~\cite{Bai}, which is performed with  COMSOL~\cite{comsol}. The dominant mode of the plasmonic dimer in the optical spectral range of interest has a quality factor $Q^{(0)}\approx 7$. We can see a clear failure of simply using $\tilde{\mathbf{f}}^{(0)}\equiv\tilde{\mathbf{f}}^{(0)}_\mu$ from Fig.~\ref{fig: QNMSingle_dimer} (b,c), where the far-field QNM $|\tilde{\mathbf{f}}^{(0)}|$, as well as the real and imaginary parts of the first-order perturbation correction $\tilde{\omega}^{(1)}=c\tilde{k}^{(1)}$ (first line of Eq.~\eqref{eq: k1_fR}) are diverging. We note here, that for the inspected structure, the results obtained from Eq.~\eqref{eq: k1_fR} are in very good agreement with the general formula, Eq.~\eqref{eq: FirstOrder_3D} (not shown). 
More rigorously, one should also account for the non-local corrections that are enforced by the continuity conditions of the electromagnetic field at the perturbation's boundary. For a spherical shape, one can take into account the non-local effects by replacing $\Delta\epsilon_{\rm p}$ by an effective permittivity change~\cite{Johnson2005,PhysRevB.79.161303}, i.e.,
$\Delta \epsilon_{\rm p} \rightarrow 3\Delta \epsilon_{\rm p} \epsilon_{\rm b}/(3 \epsilon_{\rm b}+  \Delta \epsilon_{\rm p})\equiv \Delta \epsilon_{\rm p}'$. In the presented case, we derive $\Delta \epsilon_{\rm p}'\approx 1.7$.

One heuristic approach to fix this divergence problem would be to utilize the regularization of the QNMs {\it outside} the resonator region (introduced in Section~\ref{Sec: QNMs}), 
i.e., by replacing $\tilde{\mathbf{f}}_\mu^{(0)}$ by $\tilde{\mathbf{F}}_\mu^{(0)}$ for positions $\mathbf{r}$ outside the resonator. Taking a resonant approximation, i.e., $\tilde{\mathbf{F}}_\mu(\mathbf{r},\omega)\rightarrow \tilde{\mathbf{F}}_\mu(\mathbf{r},\omega_\mu^{(0)})$, 
is needed in this case, since a $\omega$-dependent frequency change would be not in line with a modal picture. Indeed, the divergent behavior of $\tilde{\mathbf{f}}_\mu^{(0)}(\mathbf{r})$ could be circumvented in this way, as $\tilde{\mathbf{F}}_\mu^{(0)}(\mathbf{r},\omega_\mu^{(0)})$ takes on a real frequency value. However, as will be shown in Section~\ref{sec: App}, the simple replacement is not sufficient to cover all effects, that will modify the initial QNM parameters. Nevertheless, at least the spatial divergence from the usual QNM approach is prevented.

\section{Applications~\label{sec: App}}
For a more quantitative analysis, and to also use some analytical insight, we will now concentrate on one-dimensional systems.

\subsection{Single barrier problem}
\subsubsection{QNMs of a single barrier\label{subsubsec: QNMSingle}}
\begin{figure}[h]
    \centering
    \includegraphics[width=0.99\columnwidth]{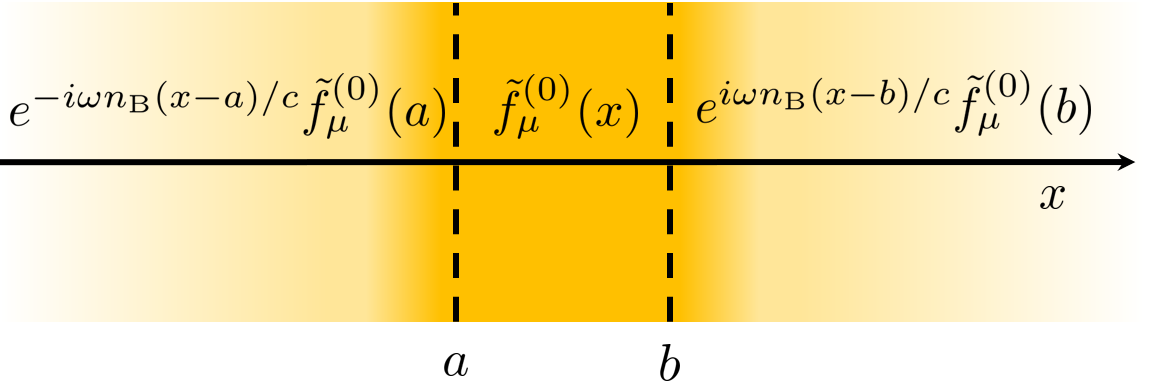}
    \caption{Schematic of a one-dimensional cavity in $x$-direction and visualization of the regularization procedure using the surface form: The QNM eigenfunctions $\tilde{f}_\mu(x)$ form a complete set inside the resonator region $x\in[a,b]$, and propagate out as plane waves through the outgoing boundary conditions, Eqs.~\eqref{eq: OutBoundCond1D}. The regularization ensures a convergent outgoing field with a real frequency $\omega$, from which one can associate a convergent regularized mode by taking $\omega\rightarrow \omega_\mu$ at the real part of the complex QNM eigenfrequencies.  
    } \label{fig: Schematic1D+Reg}
\end{figure}

We first study the problem and QNM solution of an unperturbed single cavity in the form of a single barrier with constant refractive index, spanning from $x=a$ to $x=b$, as sketched in Fig.~\ref{fig: Schematic1D+Reg}. We shall denote the length of the cavity as $L$
and the refractive index of the resonator as $n_{\rm R}$.

For the simple case of a single barrier, the QNM eigenfrequencies can be calculated via
\begin{equation}
    n_{\rm R}\tilde{k}_\mu^{(0)} L = \mu\pi + \frac{i}{2}{\rm log}\left[\frac{(n_{\rm R}-n_{\rm B})^2}{(n_{\rm R}+n_{\rm B})^2}\right],
\end{equation}
where
we made use of the outgoing boundary conditions, which read explicitly
\begin{subequations}\label{eq: OutBoundCond1D}
\begin{align}
    \partial_x \tilde{f}_\mu^{(0)}(x)\big\vert_{x\nearrow a}&=-i\tilde{\omega}_\mu n_{\rm B}/c\tilde{f}_\mu^{(0)}(a),\\
    \partial_x \tilde{f}_\mu^{(0)}(x)\big\vert_{x\searrow b}&=i\tilde{\omega}_\mu n_{\rm B}/c\tilde{f}_\mu^{(0)}(b).
\end{align}
\end{subequations}
As an aside, we observe, that the ``zero mode'' ($\mu=0$) is purely imaginary and vanishes in the lossless limit.

In the following, we choose the refractive indices $n_{\rm B}=1$ and $n_{\rm R}=2\pi$ for the background and resonator, respectively. For these parameters, we obtain
the $\mu$-independent quality factor $Q\approx 20$, resulting in an intermediate cavity finesse. We note that there are realistic examples with even lower $Q$ factor, such as the plasmonic resonator (which is commonly used in optical sensing technologies), where we expect even more drastic failure of the use of the complex QNM for positions outside the resonator region. 

The associated (normalized) QNM eigenfunctions can be derived as
\begin{equation}
    \tilde{f}_\mu^{(0)}(x) = \frac{e^{in_{\rm R}\tilde{k}_\mu^{(0)}(x-x_0)}+(-1)^{\mu}e^{-in_{\rm R}\tilde{k}_\mu^{(0)}(x-x_0)}}{\sqrt{(-1)^{\mu}2 L}n_{\rm R}},\label{eq: QNM1DBar_normalized}
\end{equation}
within the cavity region with center $x=x_0$.
\begin{figure}[h]
    \centering
    \includegraphics[width=0.99\columnwidth]{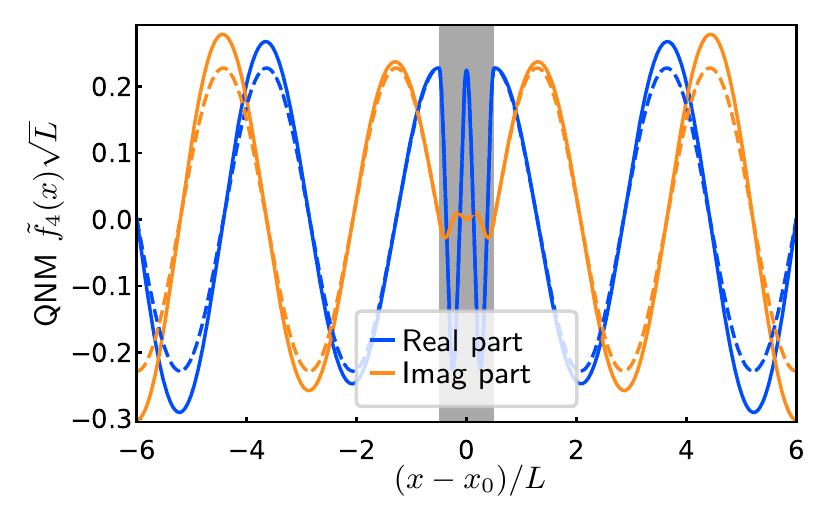}
    \caption{Real (blue - dark) and imaginary (orange - light) part of normalized QNM $\tilde{f}_\mu^{(0)}(x)$ ($\mu=4$) for the (unperturbed) single barrier cavity with center $x_0$, length $L$, background refractive index $n_{\rm B}=1$ and slab index $n_{\rm R}=2\pi$. The grey dashed area marks the cavity region. The corresponding quality factor is $Q^{(0)}\approx 20$ (and is identical for all $\mu$). The dashed lines reflects the result from a regularization outside the resonator within a pole approximation at $k={\rm Re}[\tilde{k}_4]$, i.e., $\tilde{F}_\mu^{\prime(0)}(x,\omega_\mu)$ ($\mu=4$).} \label{fig: QNMSingle_slab}
\end{figure} 

In one spatial dimension, one can formulate a generalized norm integral,
\begin{align}
    \langle\langle \tilde{f}_\mu^{(0)}|\tilde{f}_\mu^{(0)}\rangle\rangle=&\int_{a}^{b}{\rm d}x\epsilon(x)\left[\tilde{f}_\mu^{(0)}(x)\right]^2\nonumber\\
    &+i\frac{n_{\rm B }c}{2\tilde{\omega}_\mu}\left\{\left[\tilde{f}_\mu^{(0)}(a)\right]^2+\left[\tilde{f}_\mu^{(0)}(b)\right]^2\right\}\label{eq: 1DNorm},
\end{align}
where we have defined $a=x_0-L/2$, $b=x_0+L/2$.  
Indeed, for a constant permittivity $\epsilon(x)=n_{\rm R}^2$ in $[a,b]$, Eq.~\eqref{eq: 1DNorm} reduces to 
$\langle\langle \tilde{f}_\mu^{(0)}|\tilde{f}_\mu^{(0)}\rangle\rangle = 1$, when using the solution from Eq.~\eqref{eq: QNM1DBar_normalized}.
In the following, we will not specify the length $L$ of the cavity, as it simply sets the scale of the cavity problem. 

\subsubsection{Mode regularization and the local density of states}
For the one-dimensional analysis, 
the special case of Eq.~\eqref{eq: FregFEP} gives
\begin{equation}
    \tilde{F}_\mu^{\prime(0)}(x,\omega)=e^{in_{\rm B}k_0 (x-b)}\tilde{f}_\mu^{(0)}(b)\label{eq: SurFb},
\end{equation}
for $x>b$ and 
\begin{equation}
    \tilde{F}_\mu^{\prime(0)}(x,\omega)=e^{-in_{\rm B}k_0 (x-a)}\tilde{f}_\mu^{(0)}(a)\label{eq: SurFa},
\end{equation}
for $x<a$. These solutions are right- and left- propagating plane waves with continuity conditions at the boundary, i.e., $\tilde{F}_\mu^{\prime(0)}(x,\omega)\vert_{x\nearrow a}=\tilde{f}_\mu^{(0)}(a)$ and $\tilde{F}_\mu^{\prime(0)}(x,\omega)\vert_{x\searrow b}=\tilde{f}_\mu^{(0)}(b)$ independent of $\omega$. 
On the other hand, the regularized QNM from the volume integral representation, Eq.~\eqref{eq: DysonRegF2}, would simplify to
\begin{equation}
    \tilde{F}_\mu^{(0)}(x,\omega)=\frac{ik_0}{2n_{\rm B}}e^{in_{\rm B} k_0 x}\int_{a}^{b}{\rm d}s\Delta\epsilon(s)e^{-in_{\rm B} k_0 s}\tilde{f}_\mu^{(0)}(s),\label{eq: Dyson1Db}
\end{equation}
for $x>b$ and 
\begin{equation}
    \tilde{F}_\mu^{(0)}(x,\omega)=\frac{ik_0}{2n_{\rm B}}e^{-in_{\rm B} k_0 x}\int_{a}^{b}{\rm d}s\Delta\epsilon(s)e^{in_{\rm B} k_0 s}\tilde{f}_\mu^{(0)}(s),\label{eq: Dyson1Da}
\end{equation}
for $x<a$. In contrast to $\tilde{F}_\mu^{\prime(0)}(x,\omega)$ (obtained as a one-dimensional limit from the surface integral representation), $\tilde{F}_\mu^{(0)}(x,\omega)$ (obtained as a one-dimensional limit from the volume integral representation) does not fulfill the continuity conditions. However, again  we emphasize that the full electric field (including all QNMs) is still identical within both representations (cf.~App.~\ref{app: SurVolRep}). In the following we will use the surface integral representation in the few-QNM approximation.

\begin{figure}[h]
    \centering
    \includegraphics[width=0.99\columnwidth]{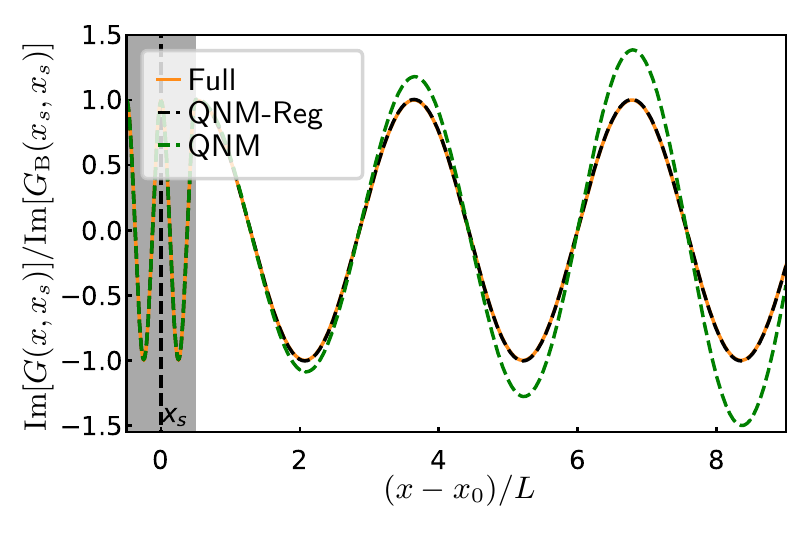}
    \caption{{Imaginary part of single barrier propagators $G(x,x_{\rm s},k={\rm Re}[\tilde{k}_4])$ normalized by ${\rm Im}[G_{\rm B}(x_{\rm s},x_{\rm s})]=1/(2n_{\rm B}{\rm Re}[\tilde{k}_4])$ with source point $x_{\rm s}=x_0$ (located in the center of the cavity) using the exact Green function Ansatz (solid orange), the single QNM approximation (green dashed) and the single QNM approximation with regularization (black dashed). Note, that $n_{\rm R}=2\pi$, which results in a quality factor of $Q\approx 20$. 
    }
}\label{fig: GF_prop_Single_Barrier_far_with_reg}
\end{figure} 
As a first test of the regularized QNM, we look at the imaginary part of the Green function, which is important for many calculation in optics and nanophotonics, e.g., the Purcell factor, which describes the enhancement/suppression of the spontaneous emission of an emitter placed close to the resonator (which uses the Green function with equal space arguments). We concentrate on a single-QNM expansion using the QNM $\mu=4$ and the parameters described in Sec.~\ref{subsubsec: QNMSingle}. As shown in Fig.~\ref{fig: GF_prop_Single_Barrier_far_with_reg}, the results using the regularized QNM are in very good agreement with the full analytical Green function solution on the basis of a plane wave expansion (without any approximation). In contrast, using the expansion with $\tilde{f}_4$ would lead to a divergent behavior, as shown by the green dashed curve in Fig.~\ref{fig: GF_prop_Single_Barrier_far_with_reg}, as expected.
This example clearly highlights the need to perform mode regularization in general, if adopting a few mode expansion approach (which is a requirement for many applications in quantum optics). One should note here, that while the imaginary part of the Green function within a single QNM approximation is in excellent agreement with the full analytical solution for the single barrier cavity, this is not the case for the real part, where one needs $N>100$ QNMs to obtain the same level of agreement.

\subsection{Double barrier problem}
\begin{figure}[h]
    \centering
    \includegraphics[width=0.99\columnwidth]{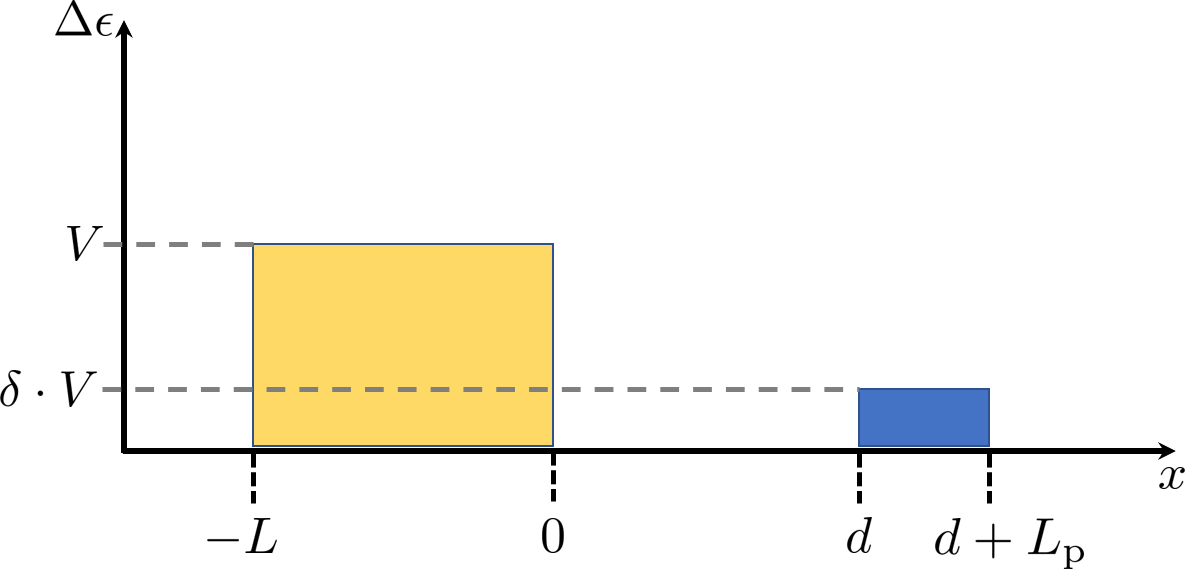}
    \caption{Permittivity difference $\Delta\epsilon(x)=\epsilon(x)-\epsilon_{\rm B}$ for the resonator-perturbation setup in one spatial dimension as function of $x$: A resonators with boundaries $x=-L,0$ and permittivity difference $\Delta\epsilon_{\rm c}(x)= V=4\pi^2-1 $ embedded in a homogeneous background medium with refractive index $n_{\rm B}=1$, including a perturbation with permittivity difference $\Delta\epsilon_{\rm p} = \delta\cdot V$, with length $L_{\rm p}$
    and distance $d$ from the cavity. 
}\label{fig: Pert_scheme}
\end{figure}

After discussing the single barrier problem in terms of of QNMs and regularized QNMs, we will now turn to the double barrier problem, as depicted in Fig.~\ref{fig: Pert_scheme}. We will concentrate on the changes that  the second barrier induces to the eigenfrequencies of single barrier cavity. This will be tackled in two different ways: (i) we will give a numerically exact solution, utilizing a transfer matrix approach, which can be used to derive the QNM poles directly or to extract them from the exact transmission (or reflection) coefficient (which obviously has no mode approximations); 
(ii)  we will apply the adapted perturbation theory to the geometry of interest, where for spatial coordinates outside the first barrier, $\tilde{f}_\mu^{(0)}$ is replaced by $\tilde{F}_\mu^{\prime(0)}$. 

\subsubsection{Direct solution from a transfer matrix approach}
Here, we discuss an exact treatment to obtain the eigenfrequencies
of the double-barrier problem, by means of a transfer matrix approach. For this, we adopt the approach of calculating the transfer matrix~\cite{PhysRevE.68.026614,10.1117/12.762652}. Note, that completeness of the QNMs in a region within the outermost discontinuities in $n(x)=\sqrt{\epsilon(x)}$ has been shown in few works. A throughout proof can be found in Ref.~\onlinecite{PhysRevE.68.026614}. Generally, in the spatial interval $[a_{j-1},a_j]$ (marking the position of the discontinuities in $n(x)$), we define the solution to the one-dimensional Helmholtz equation as forward and backward travelling waves,
\begin{equation}
    \mathcal{E}_j(x,\omega)=A_je^{ik_jx}+B_je^{-ik_jx},
\end{equation}
with $k_j = n_j\omega/c$ (and $n(x)$ is piece-wise constant) and which are connected via boundary conditions 
\begin{equation}
    \mathcal{E}_j(a_j,\omega)=\mathcal{E}_{j+1}(a_j,\omega),~\partial_x \mathcal{E}_j(a_j,\omega)=\partial_x \mathcal{E}_{j+1}(a_j,\omega).
\end{equation}

The electromagnetic fields are chosen as transverse-electric polarized fields ($E(x,t)=E_y(x,t)$).
For now, let us assume the general case of $N$ slabs, so that $j={0,1\dots,N,N+1}$ with $[0\equiv {\rm in}]$ and $[N+1\equiv {\rm out}]$ as the background indices on the left and right side, respectively. The transmission problem is defined as 
\begin{equation}
    \begin{pmatrix}
    A_{\rm in}\\B_{\rm in}
    \end{pmatrix}
    =
   \mathbf{M}_{\rm in,out}
    \cdot
    \begin{pmatrix}
     A_{\rm out}\\B_{\rm out}
    \end{pmatrix},
\end{equation}
with the {\it transfer matrix}
\begin{equation}
    \mathbf{M}_{\rm in,out}= \begin{pmatrix}
    m_{11}(\omega) & m_{12}(\omega)\\
    m_{21}(\omega)&m_{22}(\omega)
    \end{pmatrix},
\end{equation}
 which connects $A_{\rm in},B_{\rm in}$ with $A_{\rm out},B_{\rm out}$. In general, the corresponding transmission and reflection function are
\begin{equation}
    t(\omega)=\frac{A_{\rm out}}{A_{\rm in}}=\frac{1}{m_{11}(\omega)},~r(\omega)=\frac{B_{\rm in}}{A_{\rm in}}=\frac{m_{21}(\omega)}{m_{11}(\omega)}.
\end{equation}

In the case of outgoing boundary conditions, $A_{\rm in}=B_{\rm out}=0$, which implies the condition, 
\begin{equation}
    m_{11}(\omega)=0,
\end{equation}
whose complex solutions $\tilde{\omega}_\mu$ are the QNM eigenfreuqencies of the problem. Moreover, these eigenfrequencies are exactly the complex poles of the transmission and reflection functions. An explicit expression for the matrix $\mathbf{M}_{\rm in,out}$ can be obtained by subsequent multiplication of the transfer matrices from $j$ to $j+1$. For the special case of two dielectric barriers (which corresponds to three slabs), we obtain the determining equation: 
\begin{align}
    0=&\beta_{1,+}^2\beta_{2,+}^2-\beta_{1,-}^2\beta_{2,+}^2e^{2ik_{\rm R,1}L}\nonumber\\
    &+\beta_{1,-}^2\beta_{2,-}^2e^{2ik_{\rm R,1}L}e^{2ik_{\rm R,2}L_{\rm p}}-\beta_{1,+}^2\beta_{2,-}^2e^{2ik_{\rm R,2}L_{\rm p}}\nonumber\\
    &+\beta_{1,+}\beta_{1,-}\beta_{2,+}\beta_{2,-}e^{2ik_{\rm B}d}\left[e^{2ik_{\rm R,1}L}-1+e^{2ik_{\rm R,2}L_{\rm p}}\right]\nonumber\\
    &-\beta_{1,+}\beta_{1,-}\beta_{2,+}\beta_{2,-}e^{2ik_{\rm B}d}e^{2ik_{\rm R,1}L}e^{2ik_{\rm R,2}L_{\rm p}}
    ,\label{eq: DoubleBarrier_eig}
\end{align}
where $\beta_{1,\pm}=n_{\rm R,1}\pm n_{\rm B}$, $\beta_{2,\pm}=n_{\rm R,2}\pm n_{\rm B}$, and $n_{\rm R,1(2)}$ are the refractive indices of the first (second) barrier. Although this equation can only be solved analytically for certain conditions, it can be solved numerically in an exact manner. To do so, we will use the 
{\it mpmath} package in Python~\cite{mpmath}, specifically the Muller algorithm from ``mpmath.findroot''.

\subsubsection{Modification to the single barrier QNM frequencies}
\begin{figure}[hbt]
    \centering
    \includegraphics[width=0.99\columnwidth]{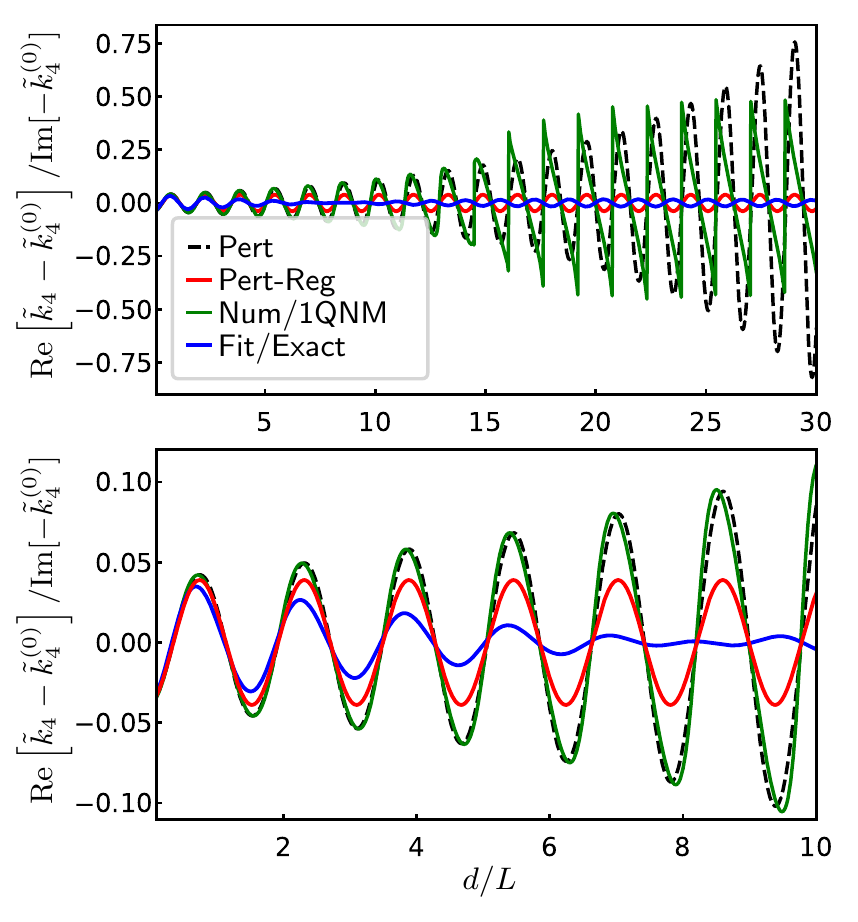}
    \caption{{Top: Real part of (normalized) eigenfrequency difference between the solution of the double barrier structure ($\tilde{k}_\mu$) and the single barrier structure ($\tilde{k}_\mu^{(0)}$) of the respective QNM with $\mu=4$ as function of distance $d$ between the two barriers, using the numerical QNM solution (solid, green), the complex QNM first-order perturbation result (black, dashed), the results from the Lorentzian fit of $t(\omega)$ (solid, blue) and the regularized QNM first-order perturbation result (solid, red).  
    The first (second) barrier has a length $L$ ($L_{\rm p}=0.1L$) and permittivity difference $\Delta\epsilon_{\rm c}=V$ ($\Delta\epsilon_{\rm p} = \delta\cdot V$ with $\delta=0.01$) with $V=4\pi^2-1$. Bottom: Enlargement (zoom in) of the top panel for smaller distances $d$. 
    }
}\label{fig: FrequencyChangeDelta001_real}
\end{figure} 

For a small height of the second barrier ($\delta\ll1$), we can apply the adapted first-order perturbation correction to obtain the frequency change
\begin{equation}
    \tilde{k}_\mu^{(1)} =-\frac{\tilde{k}_\mu^{(0)}}{2}V\frac{\left[\tilde{f}_\mu^{(0)}(0)\right]^2}{2ik_\mu^{(0)}}e^{2ik_\mu^{(0)} d}\left[ e^{2ik_\mu^{(0)} L_{\rm p}}-1\right],\label{eq: FirstOrderReg}
\end{equation}
where we have used $\tilde{F}_\mu^{\prime(0)}(x,\omega_\mu)$ instead of $\tilde{f}_\mu^{(0)}(x)$ for the evaluation of the overlap integral inside the perturbation region. It should be noted again, that we cannot strictly use the total frequency-dependent regularized QNM $\tilde{F}_\mu^{(0)}(x,\omega)$, since a $\omega$-dependent QNM frequency change would be not meaningful, and we instead impose a pole approximation around the real part $\omega_\mu$. As was shown in Ref.~\onlinecite{PhysRevB.101.205402} for a single resonator as well as for a hybrid cavity system, this approximation is excellent to capture effects for frequencies in a regime of $\sim \pm 10\gamma_\mu^{(0)}$ around the center frequency $\omega_\mu^{(0)}$.

In the following, we will compare  this correction (resulting in the modified eigenfrequency $\tilde{k}_\mu \approx \tilde{k}_\mu^{(0)}+\delta\cdot\tilde{k}_\mu^{(1)}$) to the full numerical solution obtained from Eq.~\eqref{eq: DoubleBarrier_eig}.

We first consider the real part of the frequency change as function of distance $d$, with results shown in Fig.~\ref{fig: FrequencyChangeDelta001_real}. Again, we choose an intermediate quality factor $Q\approx 20$ for the single barrier cavity by setting $n_{\rm R}=2\pi$. The height of the second barrier is $\delta V$ with $\delta=0.01$, so that one can safely assume validity of the perturbation approach. We recognize that the mode changes predicted by the full numerical pole solution are not exponentially growing, as expected. Simultaneously, perturbation theory using the naive expression involving the complex QNM $\tilde{f}_\mu$ would fail to predict the correct frequency change for larger distances, since $e^{2ik_\mu^{(0)} d}$ from Eq.~\eqref{eq: FirstOrderReg} would be replaced by  $e^{2i\tilde{k}_\mu^{(0)} d}$, leading to an exponential growth. However, when inspecting an enlargement around smaller distances (Fig.~\ref{fig: FrequencyChangeDelta001_real} (b)), one can appreciate, that the complex QNM approach agrees well with the full numerical approach. While the regularized QNM approach leads to a convergent result of the frequency change, 
it does not fully recover the results from the exact numerical approach. However, it recovers certain features of the multi-mode behavior, as will be discussed in Sec.~\ref{Sec: Multi-Mode-Eff}. Indeed, we can see that the exact single QNM also fails to describe the full solution in this regime.

\begin{figure}[h]
    \centering
    \includegraphics[width=0.99\columnwidth]{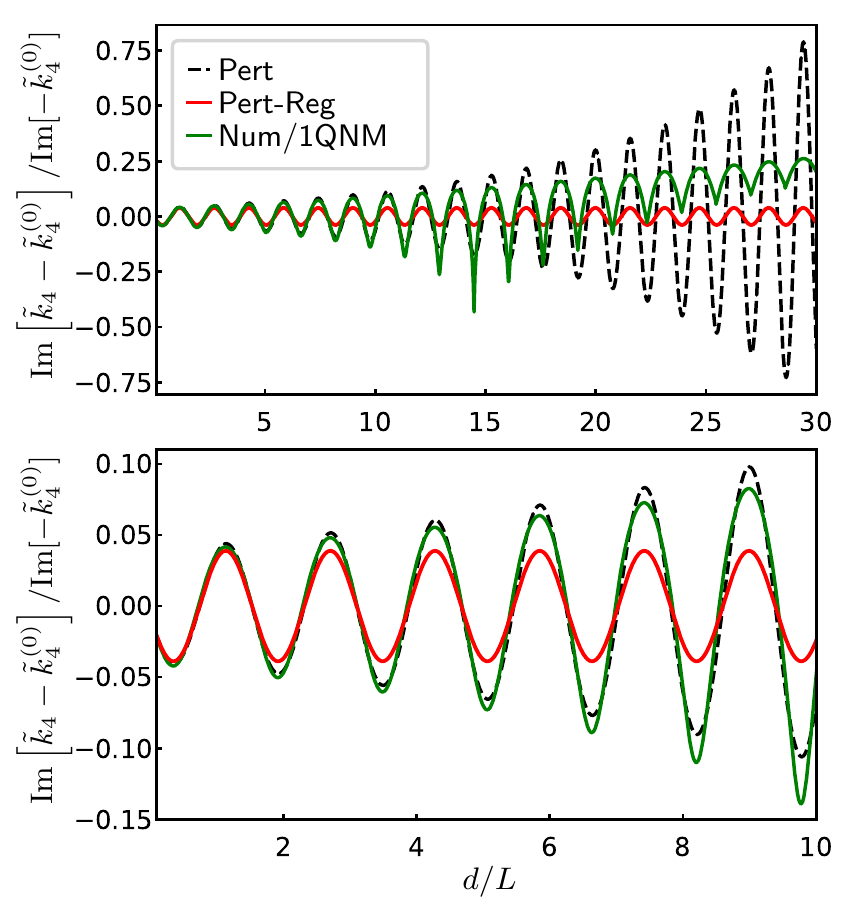}
    \caption{{Top: Imaginary part of (normalized) eigenfrequency difference between the solution of the double barrier structure ($\tilde{k}_\mu$) and the single barrier structure ($\tilde{k}_\mu^{(0)}$) of the respective QNM with $\mu=4$ as function of distance $d$ between the two barriers, 
    with the same setup as in Fig.~\ref{fig: FrequencyChangeDelta001_real}, using the numerical QNM solution (solid, green), the complex QNM perturbation result (black, dashed) and the regularized QNM perturbation result (solid, red). Bottom: Enlargement of top panel for smaller distances $d$.}
}\label{fig: FrequencyChangeDelta001_imag}
\end{figure} 

Next, we inspect the imaginary part of the frequency change, depicted in Fig.~\ref{fig: FrequencyChangeDelta001_imag}, which is related to mode dissipation. Similarly to the real part of the frequency change, the complex QNM perturbation result leads to an unphysical exponential growth of ${\rm Im}\left[\tilde{k}_\mu^{(1)}\right]$. The full numerical pole solution follows an exponential growing behavior for small distances, but it completely changes its form for $d/L>10$. While the phase relation is identical to the perturbation results, the oscillations are clearly much more complex than simple cosine and sine functions. Furthermore, the average value is detuned from $0$ to a positive value. This indicates further, that the distance from the single barrier also changes the perturbation regime. This result is  surprising and counter-intuitive, since one would expect a decreasing interaction between the two barriers for larger separations.

\subsubsection{Multi-mode effects from the full transmission coefficient\label{Sec: Multi-Mode-Eff}}
After comparing the solutions from the first-order perturbation correction to the full numerical pole solution, here we present a third method to obtain the frequency change by applying a Lorentzian fit to the transmission coefficient $t(\omega)$ around the resonance of interest (without any mode approximations). This will not only capture the changes to the individual eigenfrequencies, but also the collective effects induced by other modes. These become especially significant for lower $Q$ resonators. 

In Fig.~\ref{fig: FrequencyChangeDelta001_real}, we have already shown the corresponding results of the Lorentzian fit, which significantly differ from the predictions of the other methods. 
\begin{figure}[h]
    \centering
    \includegraphics[width=0.99\columnwidth]{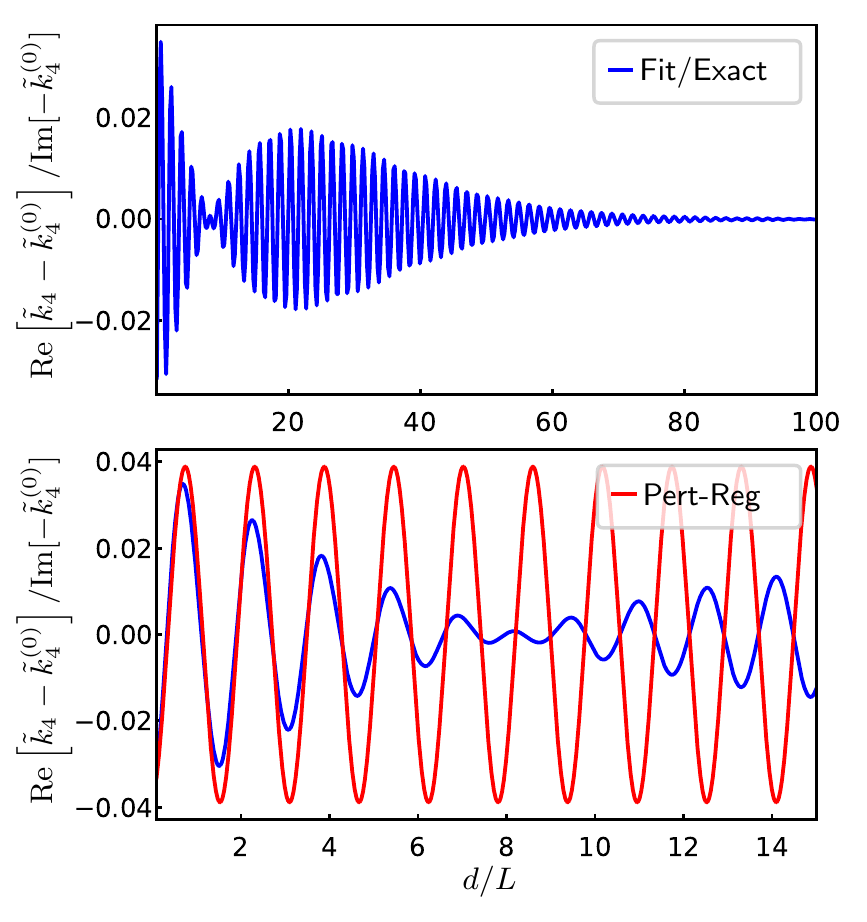}
    \caption{Real part of the eigenfrequency difference between the single barrier cavity and the double barrier cavity  (with unperturbed frequency $\tilde{k}_4^{(0)}=k_4^{(0)}-i\gamma^{(0)}_4/c$) with the same setup as in Fig.~\ref{fig: FrequencyChangeDelta001_real} using the regularized QNM perturbation result (bottom, red), and a fit result to the exact transmission coefficient (blue) as function of distance $d$.} \label{fig: Fit_vs_Reg}
\end{figure} 

In Fig.~\ref{fig: Fit_vs_Reg}, we show a more detailed analysis of the Lorentzian fit, and compare it to our proposed regularization of the QNM perturbation theory. We concentrate on the real part of the QNM eigenfrequency, i.e., the corresponding peak of the transmission coefficient around $\omega^{(0)}_4$. This is the more sensible quantity compared to the width. There are several features, that are captured by the fit results, which we summarize below: 
\begin{enumerate}
\item
A minimum appears at $d/L\sim 8$ in the fit of the exact result. 
Note that we have also observed the same feature for several parameters of the perturbation, including a change in $\delta$ and $L_{\rm p}$ (as long as $\delta<1$). Moreover, the particular position of the minimum is nearly independent of these parameters, while it is very sensitive to a change of the parameters of the single barrier cavity, such as $n_{\rm R}$.

\item
We observe a phase shift for $d/L > 8$, after passing the {\it minimum} location. To be precise, there is a phase difference of exactly $\Delta\phi = \pi/2$ compared to the QNM results (not only the regularized QNM approach).

\item Besides an additional oscillation, there is an overall damping of the eigenfrequency change, which likely reflects a multi-mode effect (fully captured by the transfer matrix results), where the changes from a single mode picture are compensated. 
\end{enumerate}

A possible reason why the complex phase oscillations are not captured by the regularization of QNMs is because one has to impose a pole approximation to $\tilde{F}(x,\omega)$, which likely  removes some interference effects, which come from other modal contributions in the exact transmission coefficient $t(\omega)$. In fact, this is not a failure of the regularization procedure itself, since its impact was recently underlined in a second quantized QNM theory, in particular for the problem of deriving a proper input-output theory for three-dimensional and absorptive cavities. However, in that case, the fast oscillating terms (namely $\exp(in_{\rm B}\omega (x-b)/c)$ for the one-dimensional analogue and $x>b$) were taken into account without any pole approximation, resulting in a time retardation. In contrast, in the 
noted 
formalism of time-independent perturbation theory, it is simply impossible to recast these $\omega$-dependent terms into a temporal shift, and thus a pole approximation is applied to the total regularized QNM function. On the other hand, we emphasize again, that from a discrete mode theory point of view, a mode quantity that depends on the continuous variable $\omega$ is not a meaningful quantity,
and we are ultimately after a convergent few mode theory. 
Consequently, one has to make some approximations, depending upon the application and problem at hand.

\section{Discussion\label{Sec: Discussions}}
The QNMs are defined as eigenfunctions of the open boundary problem, which form a complete set within the cavity region. 
The information of the ``outside'' is implicitly included in the open boundary conditions. The clear advantage over the so-called {\it modes of the universe} is that, while the modes of the universe form a continuous set of modes, the QNMs form a {\it discrete} set. The disadvantage, however, is the spatial divergence of these mode eigenfunctions. 

To elaborate further, in the usual treatment of open resonators, particularly in the field of open quantum systems, one makes an Ansatz: $H=H_0+\lambda H_I$, where $H_I$ is the coupling element with strength $\lambda$, and $H_0$ contains the energies of the isolated closed system and bath. In a second-order Born-Markov treatment of the interaction, one arrives at an effective description of the system, where the small perturbation turns into a decay rate of the system, which could in turn be regarded as a QNM imaginary part. However, in contrast to this description, the QNMs are not restricted to small imaginary parts, and the inherent mode dissipation leads to much richer properties of the QNMs (such as unconjugated forms of the normalization), which is not covered at all by the phenomenological system-bath Ansatz, as also mentioned in Ref.~\onlinecite{Leung2}. The unconjugated QNM expansion retains essential phase information, which can give rise to few mode Fano-like resonances~\cite{deLasson,KamandarDezfouli2017,PhysRevResearch.2.043290}
and unusual power flow from a point dipole~\cite{Ren2022}.

On the other hand,  regularization of the open cavity modes brings back in the properties of the background through the continuous real-valued frequency $\omega$. From a different viewpoint, the basis of the QNMs is extended by a plane wave expansion of the background Green function $\mathbf{G}_{\rm B}$, and thus again becomes continuous. Thus, in order to fully describe processes that are directly related to the ``background'', e.g., measurements at detectors outside of the cavity region, 
it seems to be inevitable to use an expansion in terms of plane waves 
in real frequency space, respectively,
even though the whole open boundary problem is solved with QNMs on their own. From a more practical point of view, one is often interested in more complicated background and cavity structures, such as a waveguide-cavity system within a photonic crystal structure. In this case, the regularized QNMs would not be formed by simple plane waves, but by waveguide Bloch modes, reflecting the periodic structure of the photonic crystal as well as the waveguide properties~\cite{Philip_CMT}. Those cases are naturally included by the proposed regularization approaches, since their appears no restriction to the background Green function, as long as their exists an analytical form, e.g., the waveguide Green function in terms of Bloch modes. 

The problem of needing additional degrees of freedom cannot only be formulated in terms of the Green function, but also in terms of the electric field. Indeed, if we inspect the surface integral representation of the regularization, Eq.~\eqref{eq: SurFb}, in the time domain, we can formulate the total electric field as
\begin{align}
    E&(x>b,t)\nonumber\\
    &=2\pi i\sum_\mu a_\mu(t=0)\tilde{f}_\mu(b)\int_{-\infty}^{\infty}{\rm d}\omega \frac{e^{-i\omega (t-n_{\rm B}[x-b]/c)}}{\tilde{\omega}_\mu-\omega}.
\end{align}

For $t>0$, we can apply the residue theorem, to obtain
\begin{align}
     E(x>b,t)&=
     \sum_\mu  a_\mu(t)\tilde{f}_\mu(b)e^{i n_{\rm B}\tilde{\omega}_\mu(x-b)/c}\nonumber\\
     &=\sum_\mu  a_\mu(t)\tilde{f}_\mu(x)\label{eq: E_reg_time},
\end{align}
for $t>n_{\rm B}[x-b]/c$ and zero otherwise. Thus, this gives another hint, that the QNMs outside the resonator cannot be formulated in terms of the system QNMs alone (i.e., the QNMs inside the resonator), but they need additional degrees of freedom. There is a clear difference when using the regularization compared to simply assuming completeness of the QNMs over all space, i.e., assuming Eq.~\eqref{eq: E_reg_time} for all $t>0$ and $x$.

The divergent behavior of the eigenfrequency change is obviously also induced by the presence of the unconjugated form of the QNM expansion, where $\tilde{f}_\mu\tilde{f}_\mu$ instead of $|\tilde{f}_\mu|^2$ appears in the first-order correction terms (cf. Eq.~\eqref{eq: FirstOrder_3D}). An interesting alternative Green function expansion by means of a conjugated form can be formulated based on a fundamental Green function relation (cf. App.~\ref{app: GreenIden}), 
\begin{align}
    {\rm Im}[G(x,y,\omega)]=n_{\rm B}\frac{c}{\omega}[&G(x,a,\omega)G^*(a,y,\omega)\nonumber\\
    &+G(x,b,\omega)G^*(b,y,\omega)],\label{eq: GreenId_1D}
\end{align}
for $a<x,y<b$, which was recognized in Ref.~\cite{2ndquanho} for the one-dimensional case with non-dispersive media. Nearly two decades later, an analogous relation for three-dimensional absorptive and leaky structures was found to be a necessary representation for QNM quantization, in which a well defined QNM Fock space can be defined~\cite{PhysRevLett.122.213901,franke2020quantized,franke2020fluctuation}. Using this knowledge from previous works, one can then utilize the Kramers-Kronig relations:
\begin{equation}
     {\rm Re}[G(x,x_0,\omega)]=\frac{1}{\pi}\mathcal{P}\int_{-\infty}^\infty{\rm d}\omega'\frac{ {\rm Im}[G(x,x_0,\omega')]}{\omega'-\omega}\label{eq: KK-1D},
\end{equation}
where $\mathcal{P}$ is the Cauchy principal value of the integral,
to obtain the corresponding real part of the Green function by means of complex contour integration techniques. This would lead to the representation (cf. App.~\ref{app: Contour1})
\begin{equation}
    G(x,x_0,\omega)=\sum_{\mu\eta}\tilde{f}_\mu(x)K_{\mu\eta}(\omega)\tilde{f}_\eta^*(x_0),\label{eq: Green1DNonDiag}
\end{equation}
where
\begin{align}
   K_{\mu\eta}(\omega)=\frac{in_{\rm B}c\tilde{\omega}_\mu}{2}\frac{\tilde{f}_\mu(a)\tilde{f}_\eta^*(a)+\tilde{f}_\mu(b)\tilde{f}_\eta^*(b)}{(\tilde{\omega}_\mu-\tilde{\omega}_\eta^*)(\omega-\tilde{\omega}_\mu)}.
\end{align}

We see a clear difference over the unconjugated form, in that a double sum over all QNMs appears. The dissipative character is now expressed in terms of off-diagonal terms, that depend on QNM functions on the cavity boundary. 

Indeed, in the NM limit ($\gamma_\mu\rightarrow 0$ and $\tilde{f}_\mu(a),\tilde{f}_\mu(b)\rightarrow 0$), the off-diagonal elements of $K_{\mu\eta}(\omega)$ tend to zero, and we are left with the diagonal elements,
\begin{align}
    K_{\mu}(\omega)=\frac{n_{\rm B}c\tilde{\omega}_\mu}{2}\frac{\tilde{f}_{\mu}(a)\tilde{f}_{\mu}^*(a)+\tilde{f}_{\mu}(b)\tilde{f}_{\mu}^*(b)}{2\gamma_{\mu}(\tilde{\omega}_{\mu}-\omega)}.
\end{align}
In the lossless limit, we can precisely derive~\cite{2ndquanho}
\begin{equation}
    n_{\rm B}c\frac{|\tilde{f}_{\mu}(a)|^2+|\tilde{f}_{\mu}(b)|^2}{2\gamma_{\mu}}\rightarrow 1,
\end{equation}
for all $\mu$, so that 
\begin{equation}
   K_{\mu\eta}(\omega)  \rightarrow \frac{\omega_\mu}{2(\omega_\mu-\omega)}\delta_{\mu\eta},
\end{equation}
and 
\begin{equation}
    G(x,x_0,\omega)\rightarrow\sum_{\mu}\frac{\omega_\mu f_\mu(x)f_\mu^*(x_0)}{2(\omega_\mu-\omega)}.
\end{equation}

If we now split the sum into $\mu=0$, $\mu>0$ and $\mu<0$, we obtain 
\begin{equation}
    G(x,x_0,\omega)\rightarrow\sum_{\mu>0}\frac{\omega_\mu^2 f_\mu(x)f_\mu^*(x_0)}{\omega_\mu^2-\omega^2},
\end{equation}
which is indeed identical to the NM expansion of the Green function. Note, here we have assumed that the NM functions have a position-independent trivial phase, so that $f_\mu(x)f_\mu^*(x_0)=f_\mu^*(x)f_\mu(x_0)$, and have used $\omega_0 \rightarrow 0$.

Note, that such a non-diagonal form of the Green function can also be derived for the more general case of three-dimensional and absorptive geometries, as shown in App.~\ref{app: 3D_NonDiagGF}.
Although the conjugated form seems to remove a potential divergent behavior, one must note, that (i) the approximation to a few QNM expansion is not equally valid in both representations, and (ii) the non-diagonal form prevents one from adapting to the perturbation theory presented in Ref.~\onlinecite{Leung2}.

\section{Conclusions\label{Sec: Conclusions}}
We have discussed  and presented different solutions to the problem of perturbation theory for open resonators, using QNMs. We showed results for the (complex) frequency change by including a perturbation outside of the resonator region, for (i) a three-dimensional plasmonic dimer structure and (ii) a one-dimensional double barrier system.
It was demonstrated that naively adopting the system QNM expansion of the Green function for positions outside the resonator yields an unphysical exponential growth of the QNM eigenfrequency change. This is induced by the dissipative nature of QNMs. We have further shown that a regularization of QNMs in terms of a function of a continuous variable for each QNM can remove such a divergent behavior of the QNMs. While neither the
few QNM or few regularized QNMs constitute an exact solution, we have shown how one can benefit from using both of these representations. Such a combination is particularly important for more complicated structures, such as coupled cavity-waveguide structures and a further theory development building upon previous works on coupled QNM frameworks~\cite{PhysRevX.11.041020,doi:10.1021/acsphotonics.1c01274} will be an important task for future investigations.

The perturbation picture could be also understood (in the quantum picture), as the problem of having a quantum dipole emitter moved away from the cavity, where one needs ({\it requires}) the regularized QNMs. To elaborate, in these applications one usually investigates the change of the emitter's linewidth and frequency through a (perturbative) second-order Born-Markov  approach (weak coupling)~\cite{PhysRevA.62.053804,PhysRevB.92.205420,PhysRevLett.122.213901}.
However, in our current investigations here, we looked at this from a different perspective, namely, how the cavity frequency and linewidth change in the presence of the emitter (perturbation) or, more generally, a perturbation object. Certainly, one would expect the same kind of fix, i.e., a regularized mode. Since the emitter has a finite linewidth, its respective radiation will also diverge in space. In a quantum perspective, their might be a deeper connection between a quantization of dissipative QNMs and a collection of emitters~\cite{clemens2003collective,clemens2004shot}, where the rigorous treatment of loss leads to an inherent coupling. In the former approach, the regularization of QNMs (in the form that is presented here) is a crucial part of the respective quantization procedure. 

\section*{Acknowledgements}
We  acknowledge funding from Queen's University,
the Canadian Foundation for Innovation, 
the Natural Sciences and Engineering Research Council of Canada, and CMC Microsystems for the provision of COMSOL Multiphysics.
We also acknowledge support from the 
Alexander von Humboldt Foundation through a Humboldt Research Award.

\appendix  
\section{Derivation of the surface representation of the regularized QNM field\label{app: SurVolRep}}
In this appendix, we proof the equivalence of Eq.~\eqref{eq: EFour_vol} and Eq.~\eqref{eq: EFour_sur}. We start with the general solution of $\boldsymbol{\mathcal{E}}(\mathbf{R},\omega)$ from the scattering problem,
\begin{align}
    \boldsymbol{\mathcal{E}}(\mathbf{R},\omega)=\int_{\mathcal{V}'}{\rm d}^3r\Delta\epsilon(\mathbf{r},\omega)\mathbf{G}_{\rm B}(\mathbf{R},\mathbf{r},\omega)\cdot\boldsymbol{\mathcal{E}}(\mathbf{r},\omega),
\end{align}
where we have chosen $\mathcal{V}'$ as a volume that is supported by $\Delta\epsilon$, i.e., where $\Delta\epsilon(\mathbf{r})\neq 0$. Next, we rewrite the above equation as 
\begin{align}
    \boldsymbol{\mathcal{E}}(\mathbf{R},\omega)=&\int_{\mathcal{V}'}{\rm d}^3r\epsilon(\mathbf{r},\omega)\mathbf{G}_{\rm B}(\mathbf{R},\mathbf{r},\omega)\cdot\boldsymbol{\mathcal{E}}(\mathbf{r},\omega)\nonumber\\
    &-\int_{\mathcal{V}'}{\rm d}^3r\epsilon_{\rm B}\mathbf{G}_{\rm B}(\mathbf{R},\mathbf{r},\omega)\cdot\boldsymbol{\mathcal{E}}(\mathbf{r},\omega).
\end{align}

Using the Helmholtz equation of $\boldsymbol{\mathcal{E}}(\mathbf{r},\omega)$ and $\mathbf{G}_{\rm B}(\mathbf{R},\mathbf{r},\omega)$, leads to
\begin{align}
    \boldsymbol{\mathcal{E}}(\mathbf{R},\omega)&=\frac{c^2}{\omega^2}\int_{\mathcal{V}'}{\rm d}^3r\mathbf{G}_{\rm B}(\mathbf{R},\mathbf{r},\omega)\cdot[\boldsymbol{\nabla}\times\boldsymbol{\nabla}\times\boldsymbol{\mathcal{E}}(\mathbf{r},\omega)]\nonumber\\
    &-\frac{c^2}{\omega^2}\int_{\mathcal{V}'}{\rm d}^3r[\boldsymbol{\nabla}\times\boldsymbol{\nabla}\times\mathbf{G}_{\rm B}(\mathbf{r},\mathbf{R},\omega)]^{\rm t}\cdot\boldsymbol{\mathcal{E}}(\mathbf{r},\omega)\nonumber\\
    &+\int_{\mathcal{V}'}{\rm d}^3r\delta(\mathbf{r}-\mathbf{R})\boldsymbol{\mathcal{E}}(\mathbf{r},\omega).
\end{align}

The last term vanishes when we choose $\mathbf{R}$, such that $\mathbf{R}\neq \mathcal{V}'$. Finally, we utilize the second dyadic-vector Green's theorem to reduce the volume integral to a surface integral over $\mathcal{S}'$:
\begin{align}
    \boldsymbol{\mathcal{E}}(\mathbf{R},\omega)&=\frac{c^2}{\omega^2}\int_{\mathcal{S}'}{\rm d}A_{\mathbf{s}} \mathbf{G}_{\rm B}(\mathbf{R},\mathbf{s},\omega)]\cdot[\mathbf{n}\times\boldsymbol{\nabla}\times\boldsymbol{\mathcal{E}}(\mathbf{s},\omega)]\nonumber\\
    &-\frac{c^2}{\omega^2}\int_{\mathcal{S}'}{\rm d}A_{\mathbf{s}}[\mathbf{n}\times\boldsymbol{\nabla}\times\mathbf{G}_{\rm B}(\mathbf{r},\mathbf{R},\omega)]^{\rm t}\cdot\boldsymbol{\mathcal{E}}(\mathbf{s},\omega).
\end{align}
The important point here is that the appearing functions and their derivatives need to be continuous in $\mathcal{V}'$, which prevents one from extending the volume $\mathcal{V}'$, in which $\Delta\epsilon$ becomes discontinuous.

If we now assume completeness of QNMs at $\mathcal{S}'$, we could reformulate the above expression as $\boldsymbol{\mathcal{E}}(\mathbf{R},\omega)=\sum_\mu a_\mu^{(0)}(\omega)\tilde{\mathbf{F}}_\mu^{\prime(0)}(\mathbf{r},\omega)$, where $\tilde{\mathbf{F}}_\mu^{\prime(0)}(\mathbf{r},\omega)$ is precisely the regularized QNM function from Eq.~\eqref{eq: FregFEP}. We emphasize, that the same derivation holds true on the level of the Green function, where the starting point would be 
\begin{align}
    \mathbf{G}^\perp(\mathbf{R},\mathbf{r})=&\mathbf{G}_{\rm B}^\perp(\mathbf{R},\mathbf{r})\nonumber\\
    &+\int{\rm d}^3s\Delta\epsilon(\mathbf{s})\mathbf{G}_{\rm B}(\mathbf{R},\mathbf{s})\cdot\mathbf{G}^\perp(\mathbf{s},\mathbf{r}),
\end{align}
with an implicit dependence on $\omega$.

\section{Proof of Green identities\label{app: GreenIden}}
In this part, we prove the Green identity from Eq.~\eqref{eq: GreenId_1D}. In one spatial dimension, the Helmholtz equation of the Green function  reduces to
\begin{equation}
    \partial_x^2G(x,x')+\frac{\omega^2}{c^2}\epsilon(x) G(x,x')=-\frac{\omega^2}{c^2}\delta(x-x'),
\end{equation}
with implicit $\omega$ dependence for the Green function and the permittivity.
The assocaited conjugated form is given by
\begin{equation}
    \partial_x^2G^*(x,x'')+\frac{\omega^2}{c^2}\epsilon^*(x) G^*(x,x'')=-\frac{\omega^2}{c^2}\delta(x-x'').
\end{equation}

For now, we assume a real permittivity $\epsilon(x)\in\mathbb{R}$ without any absorption. Note that, in principle, this is an approximation and only valid in a limited frequency range. Multiplying the first equation with $G^*(x,x'')$ and the second equation with $G(x,x')$, subtracting both resulting equations and integrating over $[a,b]$ would lead to 
\begin{align}
    2i&\frac{\omega^2}{c^2}{\rm Im}[G(x',x'')]\nonumber\\
    &=\int_a^b{\rm d}x[\partial_x^2G(x,x')]G^*(x,x'')-G(x,x')[\partial_x^2G^*(x,x'')],\label{eq: ConjCombine}
\end{align}
where we have assumed $x',x''\in[a,b]$. Utilizing partial integration techniques, we can reformulate Eq.~\eqref{eq: ConjCombine} as
\begin{align}
    2i&\frac{\omega^2}{c^2}{\rm Im}[G(x',x'')]\nonumber\\
    &=[\partial_xG(x,x')]G^*(x,x'')\vert_{a}^b-G(x,x')[\partial_xG^*(x,x'')]\vert_a^b.
\end{align}
Next, we use the outgoing boundary conditions to simplify this expression as
\begin{align}
    {\rm Im}&[G(x',x'')]\nonumber\\
    &=n_{\rm B}\frac{c}{\omega}[G(b,x')G^*(b,x'')+G(a,x')G^*(a,x'')],
\end{align}
which proves Eq.~\eqref{eq: GreenId_1D}.

\section{Complex contour integration\label{app: Contour1}}
\begin{figure}[h]
    \centering
    \includegraphics[width=0.9\columnwidth]{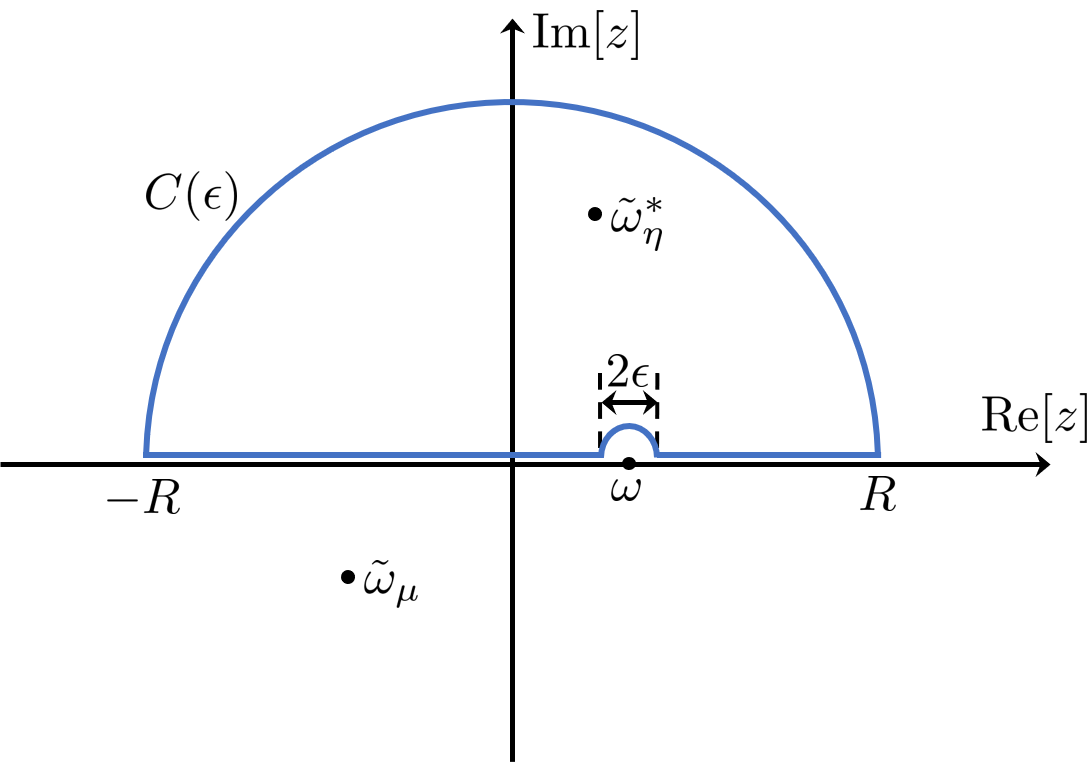}
    \caption{Visualization of the contour integration in the complex plane for the one-dimensional and non-absorptive case. The contour $C(\epsilon)$ is enclosing the set of a half circle in the upper half plane with radius $R$ (and center point $z=0$) subtracted by a smaller half circle in the upper half plane with radius $\epsilon$ (and center point $z=\omega$).
}\label{fig: Contour}
\end{figure}

In this appendix, we derive the principal value integrals that appear in the Kramers-Kronig relations, Eq.~\eqref{eq: KK-1D}, by using the non-diagonal form of the imaginary part of the Green function in one spatial dimension. The appearing frequency integral can be summarized as
\begin{equation}
    T_{\mu\eta}(\omega)=\mathcal{P}\int_{-\infty}^\infty{\rm d}\omega'\frac{\omega'}{(\omega'-\omega)(\omega'-\tilde{\omega}_\mu)(\omega'-\tilde{\omega}_\eta^*)},
\end{equation}
which can be solved analytically via complex contour integration. Specifically,  we choose a curve $C(\epsilon)$, describing a half circle with radius $R>|\omega|+\epsilon$ in the upper half plane with another smaller half circle centered at $\omega$ with radius $\epsilon$ (cf. Fig.~\ref{fig: Contour}). The contour shall contain the eigenfrequency $\tilde{\omega}_\eta^*$.

Defining the integrand as $K(\omega)$, we derive for the complex contour integration
\begin{align}
    \oint_{C(\epsilon)}&{\rm d}z K(z)=\int_{-R}^{\omega-\epsilon}{\rm d}\omega'K(\omega')+\int_{\omega+\epsilon}^{R}{\rm d}\omega'K(\omega')\nonumber\\
    +&iR^2\int_0^\pi {\rm d}\theta \frac{e^{2i\theta}}{(Re^{i\theta}-\omega)(Re^{i\theta}-\tilde{\omega}_\mu)(Re^{i\theta}-\tilde{\omega}_\eta^*)}\nonumber\\
    -&i\epsilon\int_0^\pi {\rm d}\phi \frac{e^{i\phi}(\omega+\epsilon e^{i\phi})}{\epsilon e^{i\phi}(\omega+\epsilon e^{i\phi}-\tilde{\omega}_\mu)(\omega+ \epsilon e^{i\phi}-\tilde{\omega}_\eta^*)}.
\end{align}

In the limit $R\rightarrow\infty$, the second contribution vanishes, since it scales with $1/R$. In the limit $\epsilon\rightarrow 0$, the fourth contribution remains finite and gives 
\begin{equation}
    -i\pi\frac{\omega}{(\omega-\tilde{\omega}_\mu)(\omega-\tilde{\omega}_\eta^*)}.
\end{equation}
In the limit $R\rightarrow \infty$ and $\epsilon\rightarrow 0$, the sum of the first and second contribution exactly yields the principal value of the integral of interest. Using the residue theorem, we can calculate the contour integral as 
\begin{equation}
     \oint_{C(\epsilon)}{\rm d}z K(z)=2i\pi\frac{\tilde{\omega}_\eta^*}{\tilde{\omega}_\eta^*-\tilde{\omega}_\mu}\frac{1}{\tilde{\omega}_\eta^*-\omega},
\end{equation}
so that 
\begin{equation}
    T_{\mu\eta}(\omega)=\frac{2i\pi\tilde{\omega}_\eta^*(\omega-\tilde{\omega}_\mu)+i\pi\omega(\tilde{\omega}_\mu-\tilde{\omega}_\eta^*)}{(\tilde{\omega}_\mu-\tilde{\omega}_\eta^*)(\omega-\tilde{\omega}_\eta^*)(\omega-\tilde{\omega}_\mu)}.
\end{equation}
Inserting back into the Kramers-Kronig relations and combining with the imaginary part of the Green function leads to Eq.~\eqref{eq: Green1DNonDiag}.

\section{Three-dimensional and absorptive case \label{app: 3D_NonDiagGF}}
In this part, we will present the derivation of the non-diagonal QNM Green function for the more general case of a three-dimensional and absorptive medium. This includes a generalization of the complex contour integration presented in App.~\ref{app: Contour1}.

The fundamental Green function relation in one dimension (i.e., Eq.~\eqref{eq: Green1DNonDiag}), is generalized to 
\begin{align}
    {\rm Im}&[\mathbf{G}(\mathbf{r},\mathbf{r}')]\nonumber\\
    =&\frac{c^2}{2i\omega^2}\int_{\mathcal{S}(V)}{\rm d}A_{\mathbf{s}}\Big\{[\mathbf{n}_{\mathbf{s}}\times\mathbf{G}(\mathbf{s},\mathbf{r})]^{\rm t}\cdot[\boldsymbol{\nabla}_{\mathbf{s}}\times\mathbf{G}^*(\mathbf{s},\mathbf{r}')]\nonumber\\
    &-[\boldsymbol{\nabla}_{\mathbf{s}}\times\mathbf{G}(\mathbf{s},\mathbf{r})]^{\rm t}\cdot[\mathbf{n}_{\mathbf{s}}\times\mathbf{G}^*(\mathbf{s},\mathbf{r}')]\Big\}\nonumber\\
    &+\int_V {\rm d}\mathbf{s}\epsilon_I(\mathbf{s})\mathbf{G}(\mathbf{r},\mathbf{s})\cdot\mathbf{G}^*(\mathbf{s},\mathbf{r}'),
\end{align}
for $\mathbf{r},\mathbf{r}_0\in V$, which was more rigorously derived in Ref.~\onlinecite{franke2020fluctuation}, by introducing a sequence of permittivity functions.   

Similarly to the one-dimensional case, one can utilize the Kramers-Kronig relations to obtain the corresponding real part of the Green function, ${\rm Re}[\mathbf{G}(\mathbf{r},\mathbf{r}',\omega)],
$ through 
\begin{equation}
     {\rm Re}[\mathbf{G}(\mathbf{r},\mathbf{r}',\omega)]=\frac{1}{\pi}\mathcal{P}\int_{-\infty}^\infty{\rm d}\omega'\frac{ {\rm Im}[\mathbf{G}(\mathbf{r},\mathbf{r}',\omega')]}{\omega'-\omega},\label{eq: KK_3D}
\end{equation}
where, again, $\mathcal{P}$ is the Cauchy principal value of the integral.
Here, we should note, that in contrast to the more simpler non-absorptive case, there is an additional $\omega$ kernel in the volume integral part, namely the imaginary part of the permittivity, $\epsilon_I(\mathbf{r},\omega)$. Generally, $\epsilon(\mathbf{r},\omega)$ is a causal function, i.e., an analytical function in the complex upper half plane and has the form~\cite{PhysRev.104.1760,landau2013electrodynamics} 
\begin{equation}
    \epsilon(\mathbf{r},\omega)=\epsilon_\infty(\mathbf{r})+\sum_j\frac{\sigma_j(\mathbf{r})}{\omega-\tilde{\Omega}_j},\label{eq: Mittag_Loeffler}
\end{equation}
where $\tilde{\Omega}_j$ are the complex poles of the material of interest with negative imaginary part, $\sigma_j(\mathbf{r})$ is a weighting function and $\epsilon_\infty(\mathbf{r})=\epsilon(\mathbf{r},\omega\rightarrow\infty)$ is the high-frequency limit. 

\begin{figure}[h]
    \centering
    \includegraphics[width=0.9\columnwidth]{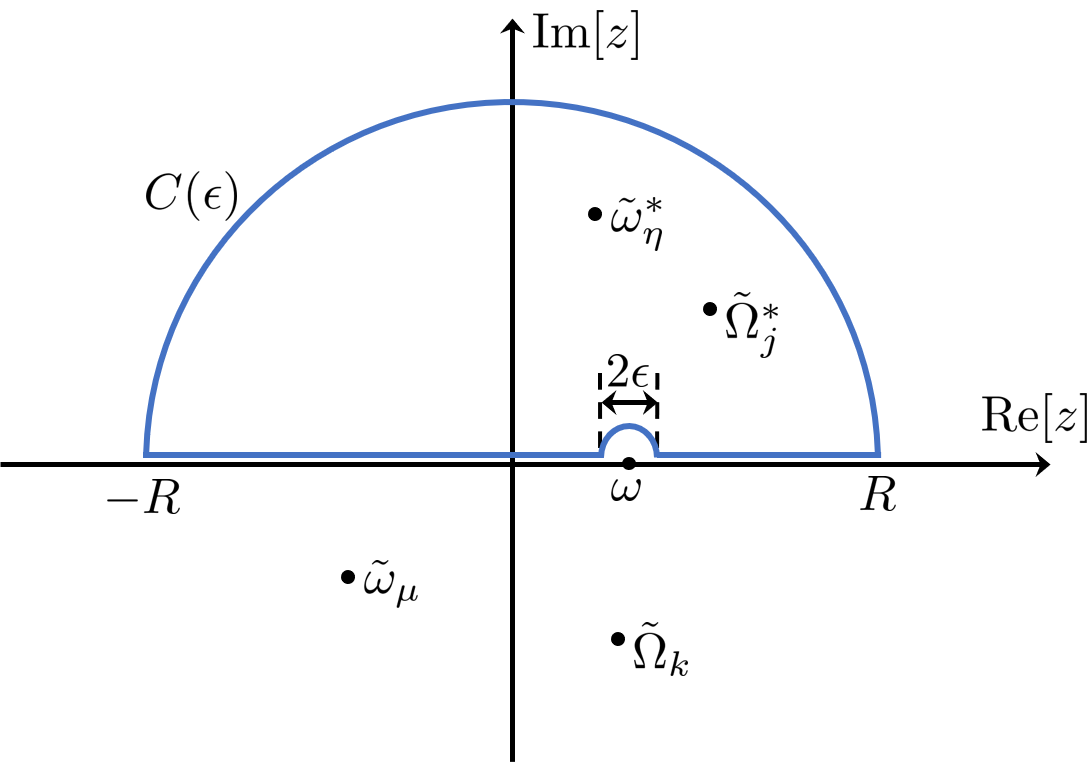}
    \caption{Visualization of the contour integration in the complex plane for the three-dimensional and absorptive case. The contour $C(\epsilon)$ is enclosing the set of a half circle in the upper half plane with radius $R$ (and center point $z=0$) subtracted by a smaller half circle in the upper half plane with radius $\epsilon$ (and center point $z=\omega$). 
}\label{fig: Contour2}
\end{figure}

First, we inspect the contribution associated with the surface integral term, where one has to solve the frequency integral
\begin{equation}
    T_{\mu\eta}^{\rm sur}(\omega)=\mathcal{P}\int_{-\infty}^\infty{\rm d}\omega'\frac{1}{(\omega'-\omega)(\omega'-\tilde{\omega}_\mu)(\omega'-\tilde{\omega}_\eta^*)},
\end{equation}
This integral can be solved analytically via complex contour integration: We choose a curve $C(\epsilon)$, describing a half circle with radius $R>|\omega|+\epsilon$ in the upper half plane with another smaller half circle centered at $\omega$ with radius $\epsilon$ (cf. Fig.~\ref{fig: Contour2}). The contour shall contain the eigenfrequency $\tilde{\omega}_\eta^*$.

Defining the integrand as $K^{\rm sur}(\omega)$, we derive for the complex contour integration
\begin{align}
    \oint_{C(\epsilon)}&{\rm d}z K^{\rm sur}(z)=\int_{-R}^{\omega-\epsilon}{\rm d}\omega'K^{\rm sur}(\omega')+\int_{\omega+\epsilon}^{R}{\rm d}\omega'K^{\rm sur}(\omega')\nonumber\\
    +&iR\int_0^\pi {\rm d}\theta \frac{e^{i\theta}}{(Re^{i\theta}-\omega)(Re^{i\theta}-\tilde{\omega}_\mu)(Re^{i\theta}-\tilde{\omega}_\eta^*)}\nonumber\\
    -&i\epsilon\int_0^\pi {\rm d}\phi \frac{e^{i\phi}}{\epsilon e^{i\phi}(\omega+\epsilon e^{i\phi}-\tilde{\omega}_\mu)(\omega+ \epsilon e^{i\phi}-\tilde{\omega}_\eta^*)},
\end{align}
where we have parametrized the half circles via $z=Re^{i\theta}$ and $z=\epsilon e^{i\phi}$, respectively.
In the limit $R\rightarrow\infty$, the third contribution vanishes, since it scales with $1/R^2$. In the limit $\epsilon\rightarrow 0$, the fourth contribution remains finite and gives 
\begin{equation}
    -i\pi\frac{1}{(\omega-\tilde{\omega}_\mu)(\omega-\tilde{\omega}_\eta^*)}.
\end{equation}
In the limit $R\rightarrow \infty$ and $\epsilon\rightarrow 0$, the first and second contribution yield exactly the principal value of the integral of interest. Using the residue theorem, we can calculate the contour integral as 
\begin{equation}
     \oint_{C(\epsilon)}{\rm d}z K^{\rm sur}(z)=2i\pi\frac{1}{(\tilde{\omega}_\eta^*-\tilde{\omega}_\mu)(\tilde{\omega}_\eta^*-\omega)},
\end{equation}
so that 
\begin{equation}
    T_{\mu\eta}^{\rm sur}(\omega)=\frac{2i\pi(\omega-\tilde{\omega}_\mu)+i\pi(\tilde{\omega}_\mu-\tilde{\omega}_\eta^*)}{(\tilde{\omega}_\mu-\tilde{\omega}_\eta^*)(\omega-\tilde{\omega}_\eta^*)(\omega-\tilde{\omega}_\mu)}.
\end{equation}

Next, we turn to the volume integral contribution, where the appearing frequency integral would read
\begin{equation}
    T_{\mu\eta}^{\rm vol}(\omega)=\frac{1}{2i}\mathcal{P}\int_{-\infty}^\infty{\rm d}\omega'\frac{[\omega']^2[\epsilon(\mathbf{r},\omega')-\epsilon^*(\mathbf{r},\omega')]}{(\omega'-\omega)(\omega'-\tilde{\omega}_\mu)(\omega'-\tilde{\omega}_\eta^*)},
\end{equation}
with an additional frequency-dependent permitivitty function in the numerator. Using the fact, that $\epsilon(\mathbf{r},\omega')$ has the form in Eq.~\eqref{eq: Mittag_Loeffler}, we derive 
\begin{align}
    T_{\mu\eta}^{\rm vol}(\omega)=&\frac{2\pi\tilde{\omega}_\eta^{*2}\chi(\mathbf{r},\tilde{\omega}_\eta^*)(\omega-\tilde{\omega}_\mu)+\pi\omega^2\chi(\mathbf{r},\omega)(\tilde{\omega}_\mu-\tilde{\omega}_\eta^*)}{2(\tilde{\omega}_\mu-\tilde{\omega}_\eta^*)(\omega-\tilde{\omega}_\eta^*)(\omega-\tilde{\omega}_\mu)}\nonumber\\
    &- \sum_j T_{\mu\eta}^{\rm vol,j}(\omega)
\end{align}
where $\chi(\mathbf{r},\omega)=\epsilon(\mathbf{r},\omega)-\epsilon_\infty(\mathbf{r})$ is the susceptibility
and 
\begin{equation}
    T_{\mu\eta}^{\rm vol,j}(\omega)=\frac{1}{2i}\mathcal{P}\int_{-\infty}^\infty{\rm d}\omega'\frac{[\omega']^2\chi^*_j(\mathbf{r},\omega')}{(\omega'-\omega)(\omega'-\tilde{\omega}_\mu)(\omega'-\tilde{\omega}_\eta^*)}.
\end{equation}
Note that there is an additional pole $\tilde{\Omega}_j^*$ in the upper complex half plane in this expression (as visualized in Fig.~\ref{fig: Contour2}), so that the residue theorem yields
\begin{align}
     \oint_{C(\epsilon)}{\rm d}z K^{\rm vol,j}(z)
     &= \pi\frac{[\tilde{\Omega}_j^{*}]^2\sigma_j(\mathbf{r})}{(\tilde{\Omega}_j^*-\tilde{\omega}_\mu)(\tilde{\Omega}_j^*-\omega)(\tilde{\Omega}_j^*-\tilde{\omega}_\eta^*)}\nonumber\\
     &+\pi\frac{[\tilde{\omega}_\eta^*]^2\chi^*_j(\mathbf{r},\tilde{\omega}_\eta^*)}{(\tilde{\omega}_\eta^*-\tilde{\omega}_\mu)(\tilde{\omega}_\eta^*-\omega)}.
\end{align}
Consequently, we obtain
\begin{align}
    T_{\mu\eta}^{\rm vol,j}(\omega)& =\pi\frac{[\tilde{\Omega}_j^{*}]^2\sigma_j(\mathbf{r})}{(\tilde{\Omega}_j^*-\tilde{\omega}_\mu)(\tilde{\Omega}_j^*-\omega)(\tilde{\Omega}_j^*-\tilde{\omega}_\eta^*)}\nonumber\\
     &+\pi\frac{[\tilde{\omega}_\eta^*]^2\chi^*_j(\mathbf{r},\tilde{\omega}_\eta^*)}{(\tilde{\omega}_\eta^*-\tilde{\omega}_\mu)(\tilde{\omega}_\eta^*-\omega)}\nonumber\\
     &+\frac{\pi}{2}\frac{\omega^2\chi^*_j(\mathbf{r},\omega)}{(\omega-\tilde{\omega}_\mu)(\omega-\tilde{\omega}_\eta^*)}.
\end{align}

Summarizing all terms, we arrive at
\begin{align}
    T_{\mu\eta}^{\rm vol}&(\omega)\nonumber\\
    =&\frac{2i\pi\tilde{\omega}_\eta^{*2}\epsilon_I(\mathbf{r},\tilde{\omega}_\eta^*)(\omega-\tilde{\omega}_\mu)+i\pi\omega^2\epsilon_I(\mathbf{r},\omega)(\tilde{\omega}_\mu-\tilde{\omega}_\eta^*)}{(\tilde{\omega}_\mu-\tilde{\omega}_\eta^*)(\omega-\tilde{\omega}_\eta^*)(\omega-\tilde{\omega}_\mu)}\nonumber\\
    &+ \pi\frac{[\tilde{\Omega}_j^{*}]^2\sigma_j(\mathbf{r})}{(\tilde{\Omega}_j^*-\tilde{\omega}_\mu)(\tilde{\Omega}_j^*-\omega)(\tilde{\Omega}_j^*-\tilde{\omega}_\eta^*)}.
\end{align}
Finally, inserting back into the Kramers-Kronig relation \eqref{eq: KK_3D}
yields
\begin{equation}
    \mathbf{G}(\mathbf{r},\mathbf{r}',\omega)=\sum_{\mu\eta}\tilde{\mathbf{f}}_\mu(\mathbf{r})K_{\mu\eta}(\omega)\tilde{\mathbf{f}}_\eta^*(\mathbf{r}')\label{eq: GF_non_diag_3D},
\end{equation}
where $K_{\mu\eta}(\omega)=K_{\mu\eta}^{\rm sur}(\omega)+K_{\mu\eta}^{\rm vol}(\omega)$. Here, 
\begin{align}
   K_{\mu\eta}^{\rm sur}(\omega)
   =\frac{c^2\int_{\mathcal{S}(V)}{\rm d}A_{\mathbf{s}}\Big\{I_{\mu\eta}(\mathbf{s})-I_{\eta\mu}^*(\mathbf{s})\Big\}}{4(\tilde{\omega}_\mu-\tilde{\omega}_\eta^*)(\omega-\tilde{\omega}_\mu)},
\end{align}
with $I_{\mu\eta}=[\mathbf{n}_{\mathbf{s}}\times\tilde{\mathbf{f}}_\mu(\mathbf{s})]^{\rm t}\cdot[\boldsymbol{\nabla}_{\mathbf{s}}\times\tilde{\mathbf{f}}_\eta^*(\mathbf{s})]$ being the surface contribution.

Furthermore,  
\begin{align}
   K_{\mu\eta}^{\rm vol}(\omega)
   =\frac{\int_{V}{\rm d}\mathbf{s} R_{\mu\eta}(\mathbf{s},\omega)\tilde{\mathbf{f}}_\mu(\mathbf{s})\cdot\tilde{\mathbf{f}}_\eta^*(\mathbf{s})}{4(\tilde{\omega}_\mu-\tilde{\omega}_\eta^*)(\omega-\tilde{\omega}_\mu)}+M_{\mu\eta}(\omega)
\end{align}
with 
\begin{align}
    R_{\mu\eta}(\mathbf{s},\omega)=\frac{2i\tilde{\omega}_\eta^{*2}\epsilon_I(\mathbf{r},\tilde{\omega}_\eta^*)(\omega-\tilde{\omega}_\mu)}{\omega-\tilde{\omega}_\eta^*}\nonumber\\
    +\frac{2i\omega^2\epsilon_I(\mathbf{r},\omega)(\tilde{\omega}_\mu-\tilde{\omega}_\eta^*)}{\omega-\tilde{\omega}_\eta^*}
\end{align}
and 
\begin{equation}
    M_{\mu\eta}(\omega)=\sum_j\frac{[\tilde{\Omega}_j^{*}]^2\int_{V}{\rm d}\mathbf{s}\sigma_j(\mathbf{r})\tilde{\mathbf{f}}_\mu(\mathbf{s})\cdot\tilde{\mathbf{f}}_\eta^*(\mathbf{s})}{4(\tilde{\Omega}_j^*-\tilde{\omega}_\mu)(\tilde{\Omega}_j^*-\omega)(\tilde{\Omega}_j^*-\tilde{\omega}_\eta^*)},
\end{equation}
originates from the volume part of the Green identity.

Taking the subsequent limits of $\sigma_j\rightarrow 0$ and $\gamma_\mu\rightarrow 0$ (and vanishing QNM functions on the cavity boundary and in the background region), we 
again recover the usual NM expansion, similar to the one-dimensional case. A formal proof of that limit in combination with a specific inner product of QNMs can be found in Ref.~\onlinecite{franke2020fluctuation}.

\end{document}